\documentclass[fleqn,usenatbib]{mnras}

\usepackage{newtxtext,newtxmath}

\usepackage[T1]{fontenc}

\DeclareRobustCommand{\VAN}[3]{#2}
\let\VANthebibliography\thebibliography
\def\thebibliography{\DeclareRobustCommand{\VAN}[3]{##3}\VANthebibliography}

\usepackage{graphicx}	
\usepackage{amsmath}	
\usepackage{subcaption}

\title[Towards modeling Ghostly DLAs]{
Towards modeling Ghostly DLAs
}

\author[B. Laloux 
]{Brivael Laloux$^{1,\,2}$,
Patrick Petitjean$^{3}$,
Camille No\^us$^4$
\\
$^{1}$Centre for Extragalactic Astronomy, Department of Physics, Durham
University, Durham, DH1 3LE, UK\\
$^{2}$National Observatory of Athens, V. Paulou \& I. Metaxa, 11532, Greece\\
$^{3}$Institut d'Astrophysique de Paris, Sorbonne Universit\'es and CNRS, 98bis Boulevard Arago, 75014 Paris, France\\
$^4$Laboratoire Cogitamus, 98bis, Boulevard Arago, 75014 Paris, France\\
}

\date{Accepted XXX. Received YYY; in original form ZZZ}

\pubyear{2020}

\begin{document}
\label{firstpage}
\pagerange{\pageref{firstpage}--\pageref{lastpage}}
\maketitle

\begin{abstract}
We use simple models of the spatial structure of the quasar broad line region (BLR) to investigate the properties of so-called \textit{ghostly} damped Lyman-$\alpha$ (DLA) systems detected in SDSS data. These absorbers are characterized by the presence of strong metal lines but no H~{\sc i} Lyman-$\alpha$ trough is seen in the quasar spectrum indicating that, although the region emitting the quasar continuum is covered by an absorbing cloud, the BLR is only partially covered. 
One of the models has a spherical geometry, another one is the combination of two wind flows whereas the third model is a Keplerian disk. 
The models can reproduce the typical shape of the quasar Lyman-$\alpha$ emission 
and different \textit{ghostly} configurations. We show that the DLA H~{\sc i} column density can be recovered precisely independently of the BLR model used.
The size of the absorbing cloud and its distance to the centre of the AGN are correlated. However it may be possible to disentangle the two using an independent estimate of the radius from the determination of the particle density. Comparison of the model outputs with SDSS data shows that the wind and disk models are more versatile than the spherical one and can be more easily adapted to the observations. For all the systems we derive log~$N$(H~{\sc i})(cm$^{-2}$)~$>$~20.5. With higher quality data it may be possible to distinguish between the models. 
\end{abstract}

\begin{keywords}
quasars:  absorption lines — quasars:  emission lines
\end{keywords}



\section{Introduction}\label{section: Intro}

One of the most challenging issues in quasar physics is to understand how black-holes are fed with infalling gas. The infall of gas onto the host-galaxy occurs preferentially through cold streams along the filaments of the cosmic web \citep{2012MNRAS.421.2809V}. So far, no direct robust observational evidence has been found for the existence of this infalling gas \citep{2006A&A...459..717C, O_Sullivan_2020}. 
Instead, outflows driven by active galactic nuclei (AGN) are ubiquitously observed as blue-shifted absorption features in quasar spectra at any redshift
\citep{Rankine_2020}.

The infalling gas must be compressed when reaching the disk of the galaxy and the compressed gas could give rise to a DLA in the quasar spectrum.
Since the DLA and the background quasar are located at almost the same redshift, the DLA can act as a natural coronagraph, blocking the quasar blazing radiation in Lyman-$\alpha$. This can allow us, depending on the dimension of these so-called \emph{eclipsing} DLAs, to detect fainter emission from star-forming regions in the host galaxy and the extended quasar halo and/or to observe the narrow line region (NLR) of the AGN. The leaked emission from  these regions can be detected as a narrow Lyman-$\alpha$ emission in the DLA trough \citep{2009ApJ...693L..49H, 2013A&A...558A.111F}

If the absorbing cloud gets denser, smaller and closer to the quasar then we expect the narrow emission line in the DLA trough to increase in strength. In extreme cases where the hydrogen density is very high (i.e. $n_{\rm HI}$\,$>$\,1000\,cm$^{-3}$) and the cloud size is smaller than the size of the quasar BLR, the leaked broad Lyman-$\alpha$ emission from the BLR can fill the DLA trough completely and the DLA absorption profile is therefore not seen in the spectrum \citep{Fathivavsari_2017}.
This is why these DLAs are called \textit{ghostly}-DLAs. The characterization of this kind of systems is extremely important to understand the details of how the neutral gas ends up at this position in such a harsh environment.

Conversely, the fact that a cloud smaller than the typical BLR size only partially covers the emission can constrain the spatial structure of the emission. These systems are potentially a powerful tool to study the structure of the BLR.

The BLR is thought to be composed of approximately virialised gas in the vicinity of the black hole \citep{Netzer_2008}. From this idea, it is possible to derive the typical size of the emission by performing reverberation mapping analysis \citep{2019ApJ...883L..14S}.
These studies reveal an expected correlation between the BLR size and the central luminosity \citep{2013ApJ...767..149B}. However, it is also possible that at least part of the broad emission lines are produced by outflowing material launched from near the accretion disc.
This is most strikingly suggested by observations of Broad Absorption Lines (BALs) in about 20\% of quasars and the link between the emission lines and BALs has been studied in details  \citep{2020MNRAS.492.5540M}. Observationally, reverberation mapping of the H$\beta$ emission of quasars at low-redshift has resulted in constraining the geometry and kinematics of the region emitting this line. \cite{2017ApJ...851...21G} found these emission regions to be thick disks that are close to face-on to the observer with kinematics that are well-described by either elliptical orbits or inflowing gas. Time lags as a function of the velocity across the H$\beta$ emission line profile have been measured in a number of AGNs.
Various kinematic signatures have been found in the different objects; these kinematic signatures are mostly virialized motions and inflows but also outflows \citep{2018MNRAS.476..943H, 2019A&A...630A..94G, 2016ApJ...820...27D}. 

These studies have been complemented by analysis of microlensing amplification of quasar continua and emission lines. Microlensing-induced line profile deformations analysis can constrain the BLR size, geometry and kinematics \citep{1990A&A...237...42S}.
Comparisons with models reveal that  strong microlensing effects put important constraints on the size of the BLR \citep{Braidant_2017}. Comparisons with observations show that flattened geometries (Keplerian disk and equatorial wind) can more easily reproduce the observed line profile deformations than a biconical polar wind \citep{2019A&A...629A..43H}.

In this paper, we construct simple models of the BLR, partially covered by an absorbing cloud, and  use these models to characterize and fit observations of quasar spectra bearing \textit{ghostly}-DLAs.
In these spectra, although a DLA cloud is present in front of the quasar, no Lyman-$\alpha$ trough is detected whereas a Lyman-$\beta$ trough, when redshifted in the observed wavelength window, is clearly seen. 
We use the fact that only part of the BLR is covered to investigate whether it will be possible to differentiate between models and to constrain some properties of the BLR and of the absorbing cloud. 
An important starting point of our models is that we require them to reproduce the typical spectrum of a bright high-redshift quasar represented by a quasar template.

We describe the models in Section 2, explore how the models can produce \textit{ghostly}-DLAs in Section 3, investigate the use of the models by fitting mock spectra in Sections 4 and 5, fit real SDSS data in Section 6 and draw conclusions in Section 7.

\section{Modelling the quasar spectrum} \label{section: modelling}

In the following, we model the quasar spectrum in the Lyman-$\alpha$ and Lyman-$\beta$ emission regions. 
The quasar is described as a central point-like source emitting a power-law continuum surrounded by a broad line region described as a distribution of clouds with particular spatial and kinematic structures (see below) and a more extended narrow-line region (NLR). Each cloud of the BLR is supposed to emit the same amount of Lyman-$\alpha$ photons.  The rest-frame emission of each cloud is modelled as a Gaussian emission line of width FWHM~=~50~km~s$^{-1}$. 
The stratification of the BLR is defined by the density of clouds through the structure. 
Transfer of Lyman-$\alpha$ photons is not considered which means that we assume the covering factor of the BLR clouds to be small. The BLR emission line is the superposition of the individual emissions of the clouds after taking into account their velocities.  
We add a narrow emission line to the spectrum corresponding to the NLR  emission. This region will be assumed not to be covered by the absorbing cloud. 
As described below, we will use three different geometrical models of the BLR: a spherical model, a wind model and a Keplerian disk model. 

The typical radius of the high-redshift quasar BLR is of the order of one parsec. However our models do not depend on the exact radius of the BLR and in the following, radial dimensions in the BLR or in the cloud will be defined as the unit free ratio $r\equiv r_{\rm 0}/r_{\rm BLR}$ where $r_{\rm 0}$ is the real radial dimension and $r_{\rm BLR}$ is the BLR radius, both in pc units.

\begin{figure}
    \centering
    \includegraphics[trim = 2.cm 0cm 2.5cm 1.5cm, clip,width=0.5\textwidth]{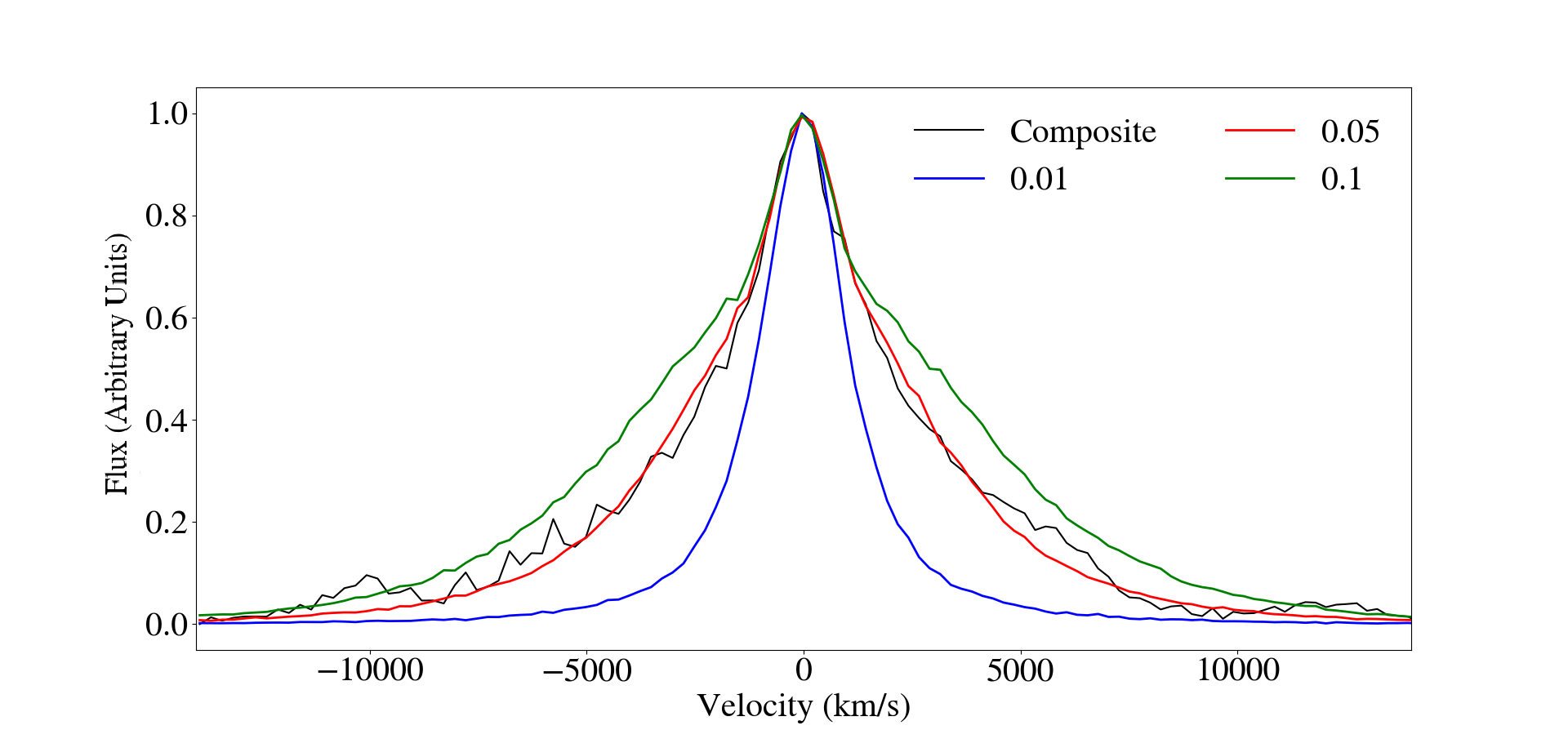}
    \caption{Comparison of spectra derived from the spherical model with the quasar composite spectrum after subtraction of the quasar continuum and the N~{\sc v} emission and represented by the black line. The blue, red and green curves correspond to $\sigma_0$~=~9,000~km~s$^{-1}$ and $r_{\rm min}$~=~0.01, 0.05 and 0.1, respectively.}
    \label{fig:dif_rmin_composite}
\end{figure}

To adjust the parameters of the models, we fit their outputs to a composite quasar spectrum obtained using 2200 quasar spectra of the Sloan Digital Sky Survey \citep{composite}.
Since we are interested in the quasar Lyman-$\alpha$ emission,  we subtract the N~{\sc v} emission from the template. For this, we fit the composite spectrum with two sets of two Gaussians representing the Lyman-$\alpha$ and N~{\sc v} emissions. 
The widths of the Gaussian functions are the same for the two emissions.
We then remove the contribution of the N~{\sc v} $\lambda$1240 emission line to obtain the typical Lyman-$\alpha$ quasar emission (represented by the black line on e.g. Fig.~\ref{fig:dif_rmin_composite}).

To model the spectrum of a \textit{ghostly}-DLA, we will add an absorbing cloud on top of the continuum and the BLR emission, the narrow line region staying uncovered.

\begin{figure}
    \centering
    \includegraphics[trim = 2cm 0.2cm 3.2cm 1.8cm, clip,width=0.45\textwidth]{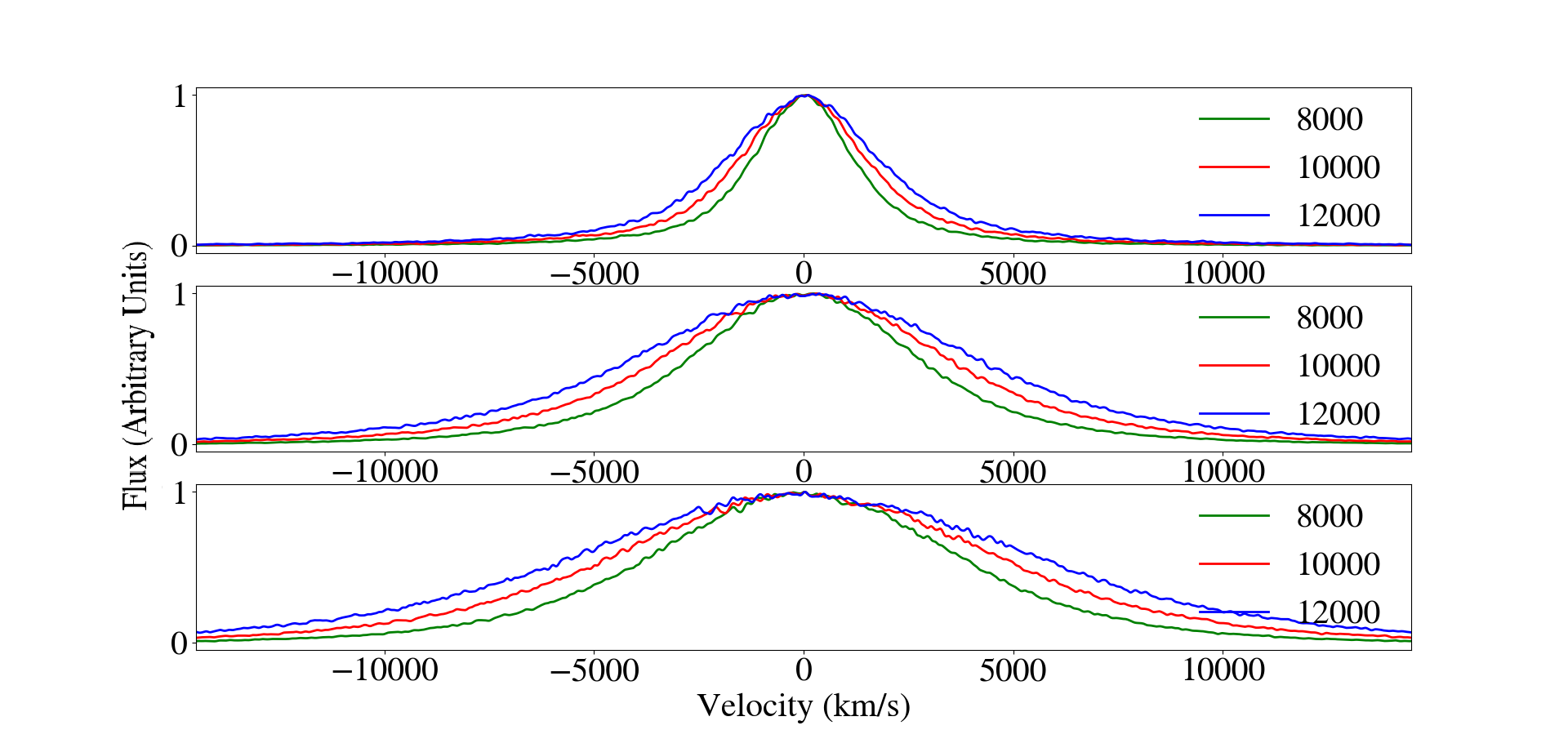}
    \caption{Modelled Lyman-$\alpha$ emission spectra of the spherical BLR model as a function of the maximum standard deviation (8,000, 10,000 and 
    12,000~km~s$^{-1}$ for the green, red and blue curves respectively)
    for different inner radius, $r_{\rm min}=0.01$, 0.05 and 0.1, for the top to bottom panels, respectively.}
    \label{fig:dif_rmin}
\end{figure}

\subsection{Spherical model}

\begin{figure}
    \centering
    \begin{subfigure}[b]{0.5\textwidth}
        \includegraphics[trim = 4.1cm 1cm 3cm 1cm, clip,width=\textwidth]{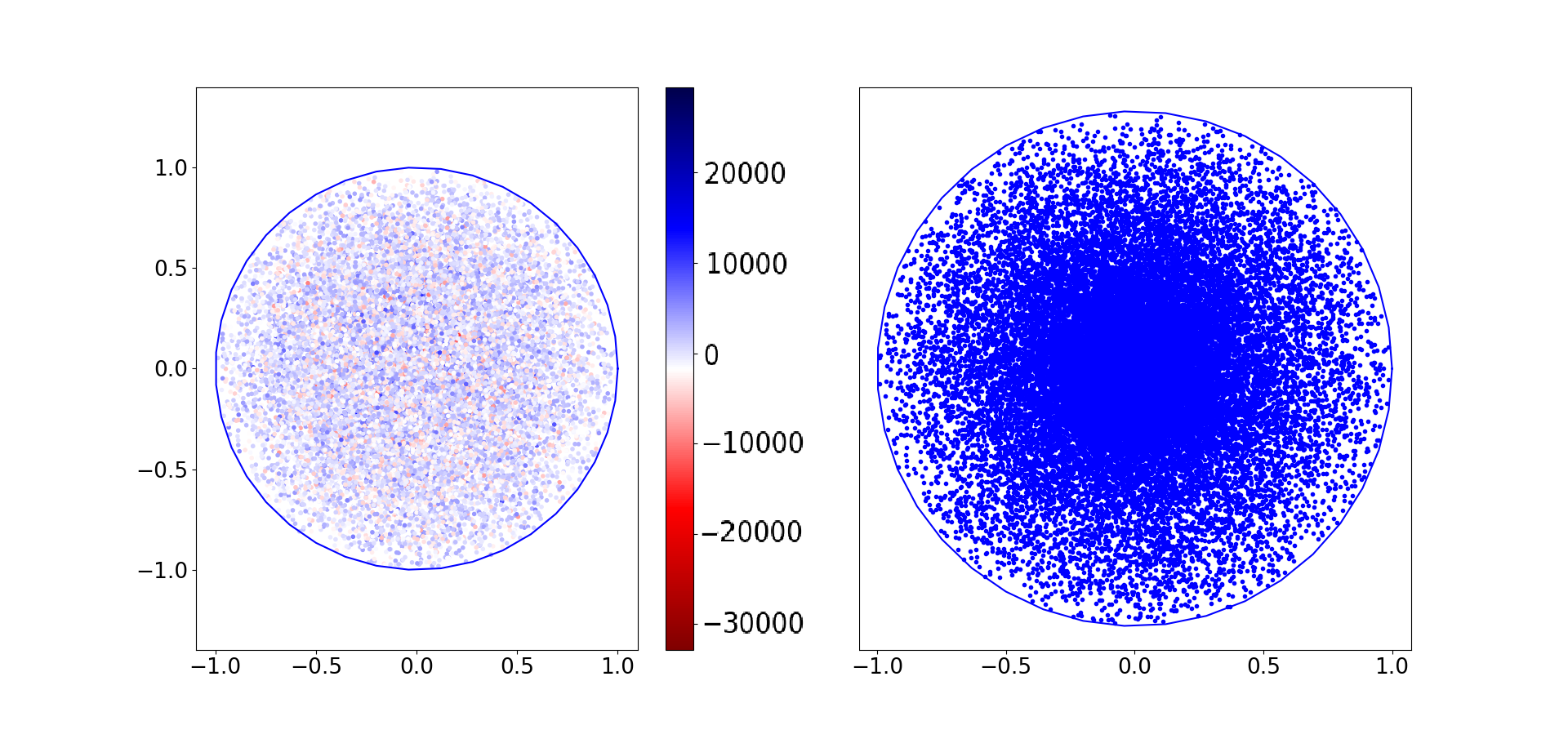}
        \caption{}
        \label{fig:dispo_spherical}
    \end{subfigure}
    \begin{subfigure}[b]{0.5\textwidth}
        \includegraphics[trim = 2cm 0.cm 2.5cm 1.5cm, clip,width=\textwidth]{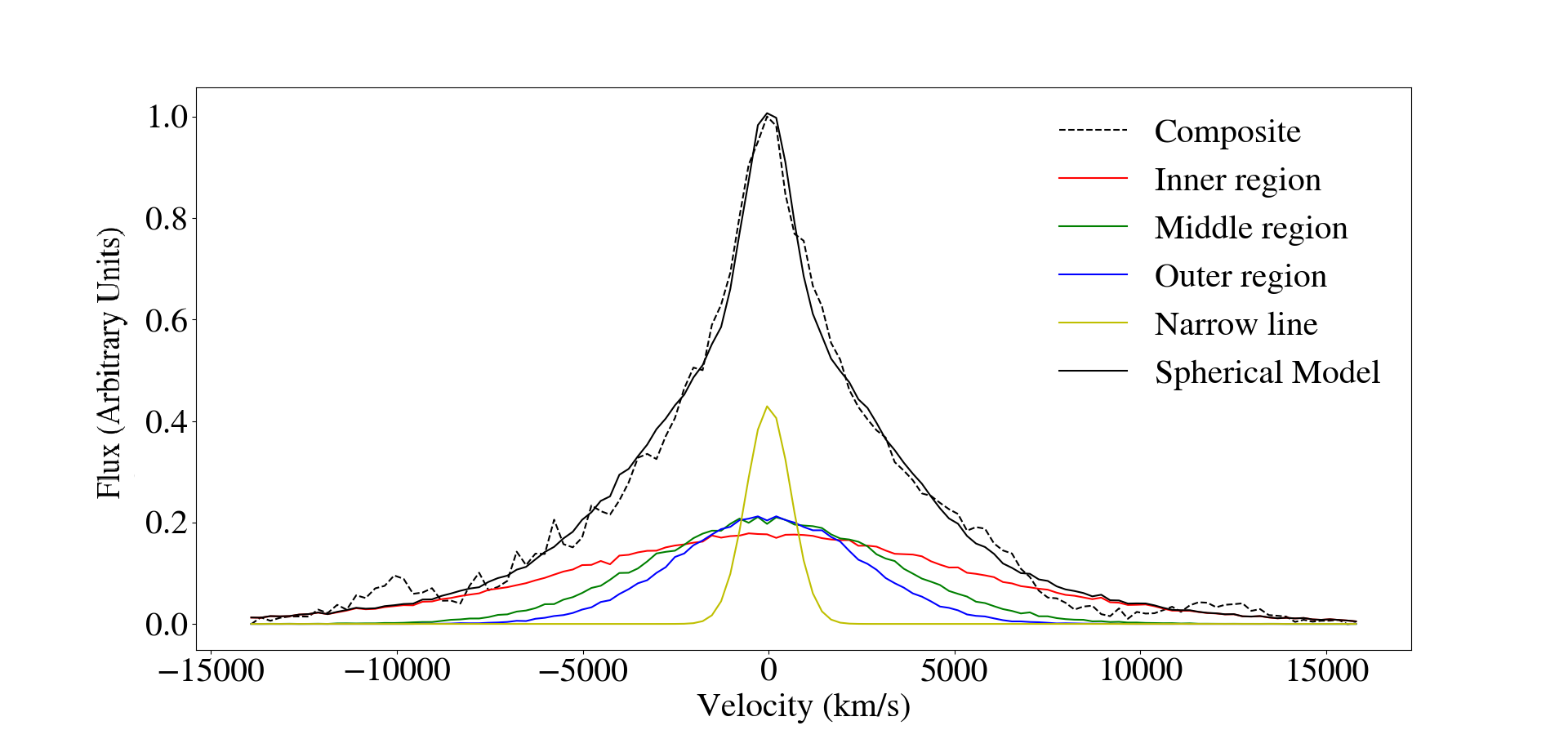}
        \caption{}
        \label{fig:spectrum_spherical}
    \end{subfigure}
    \caption{Layout of the spherical BLR model (a) and its corresponding spectrum (b). In the top left-hand side panel, the color  scale corresponds to the velocity (in km~s$^{-1}$) of the clouds relative to the observer. 
    In the right-hand side panel, the blue dots illustrate the variation of the density of clouds in the BLR. In panel (b), the composite quasar spectrum is fitted with the spherical model with $\sigma_0 =$ 10000~km~s$^{-1}$, a BLR/NLR strength ratio of 2.33 and the width of the narrow line of 500~km~s$^{-1}$. The contributions of different regions of the BLR (inner for 0~$<$~$r$~$<$~0.33, middle for 0.33~$<$~$r$~ $<$~0.66 and outer for 0.66~$<$~$r$~$<$1) are singled out and shown as red, green and blue lines respectively.}
    \label{dispo_spectre_sphere}
\end{figure}

In this model, the distribution of point-like clouds around the quasar is spherical. 
The density of clouds depends on the distance to the centre $r$ and is given by the following density profile: 
\begin{equation}
    n=n_0 \,\left(\frac{r}{r_{\rm min}}\right)^{-\alpha}
    \label{eq: density}
\end{equation}
where $n_0$ is the number of emitting clouds at the internal radius $r_{\rm min}$, the sphere being empty from $r$~=~0 to $r$~=~$r_{\rm min}$. We use $\alpha$~=~0.5. 

The velocity of each cloud relative to the observer is random and follows a Gaussian probability function of dispersion $\sigma$ \citep{1996ApJ...463..144D}. 
This dispersion depends on the distance to the centre and behaves according to Keplerian laws:
\begin{equation}
    \sigma=\sigma_0 \,\left(\frac{r}{r_{\rm min}}\right)^{-0.5}
    \label{eq: v sphere}
\end{equation}
where $\sigma_0$ is the maximum dispersion.

To fill the BLR with clouds, the sphere is divided into 1100 layers in which emitting clouds are randomly distributed one by one until reaching the desired density in the layer. The first inner layer has 1100 clouds. The total number of clouds in the BLR is 356,400. These numbers are chosen so that the resulting spectrum is smooth
enough keeping the computing time reasonable.

Given the above velocity law, it happens that amongst the two parameters which determine the width of the BLR Lyman-$\alpha$ emission, $r_{\rm min}$ and $\sigma_0$, $r_{\rm min}$ is the most important. On Fig.~\ref{fig:dif_rmin}, we represent the Lyman-$\alpha$ emission of the BLR for three different values of $r_{\rm min}$ and $\sigma_0$. 

On  Fig.~\ref{fig:dif_rmin_composite}, we fit the composite spectrum (after subtraction of the quasar continuum) with a modelled quasar spectrum built from the addition of the BLR and NLR emissions.

We find that the spectrum is reasonably well reproduced with 0.03~$\leq$~$r_{\rm min}$~$\leq$~0.05 and 8500~$\leq$~$\sigma_0$~$\leq$~10000~km~s$^{-1}$. The width of the NLR emission is in the range 400~$-$~700~km~s$^{-1}$. 
For the rest of the paper, we will fix $r_{\rm min} = 0.05$ and consider $\sigma_0$ and the width of the NLR as free parameters.

The spatial layout of the emitting clouds in the BLR and the corresponding spectrum for the spherical model are shown in Fig.~\ref{fig:dispo_spherical} and Fig.~\ref{fig:spectrum_spherical}, respectively. The dots are colored according to their velocity relative to the observer in the left-hand side panel. 
In the right-hand side panel, the dots are plotted with the same color to better illustrate the variation of the density within the BLR. 

We then single out three regions as a function of their distance to the centre, the inner region from the centre to $r$~=~0.33, the intermediate region from $r$~=~0.33 to 0.66, and the outer region from $r$~=~0.66 to 1. Their respective contributions to the spectrum are represented on Fig.~\ref{fig:spectrum_spherical} by a red, green and blue line, respectively. Dots located near the centre have a larger dispersion in velocities (see Eq.~\ref{eq: v sphere}) than the one located further away, this is why the wings of the spectrum are only produced by the emitting clouds located in the center.

\begin{figure}
\centering
\includegraphics[trim = 0.5cm 0cm 5.5cm 0cm, clip,width=0.45\textwidth]{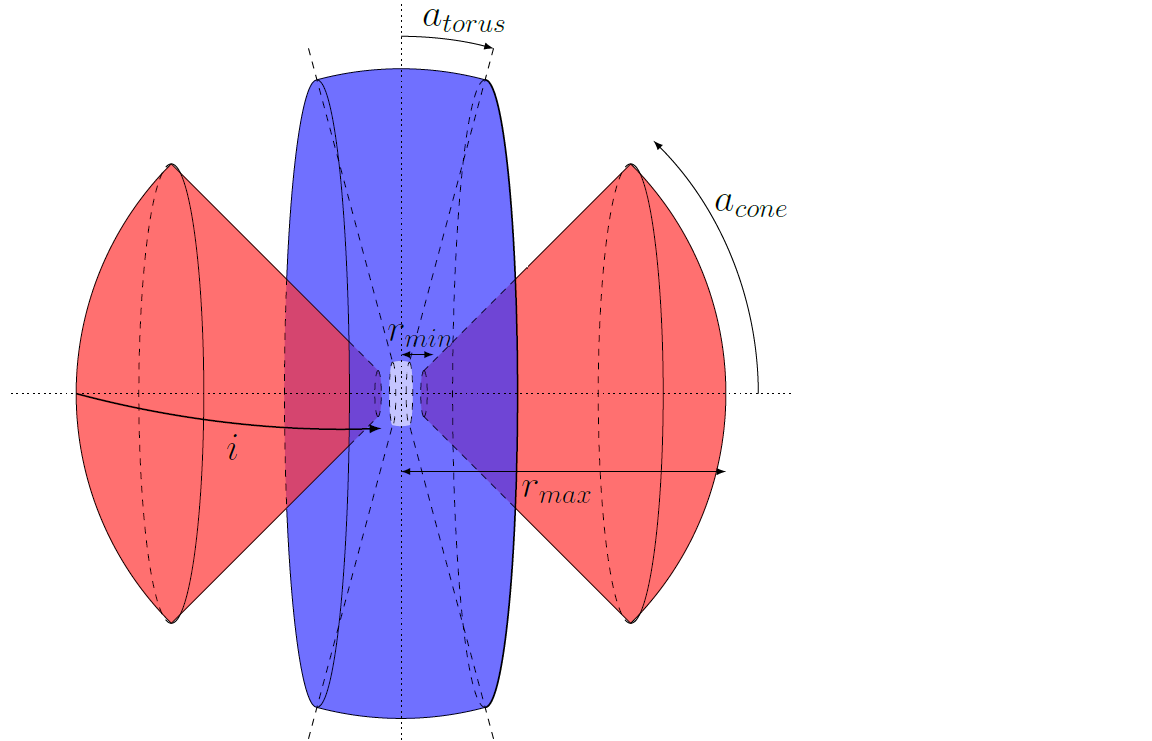}
\caption{Sketch of the wind model with an inclination  of the model axis relative to the line of sight to the observer, $i\sim85$°. In red, the polar wind with an opening angle $a_{\rm cone}=45$°. In blue, the equatorial wind with an angle $a_{\rm torus}=15$°. Both winds have an internal radius $r_{\rm min}=0.1$ and an external 
radius $r_{\rm max}~=~1$.} 
\label{schema}
\end{figure}

\subsection{Wind model}
\begin{figure*}
    \includegraphics[trim = 4cm 1cm 3cm 2cm, clip,width=\textwidth]{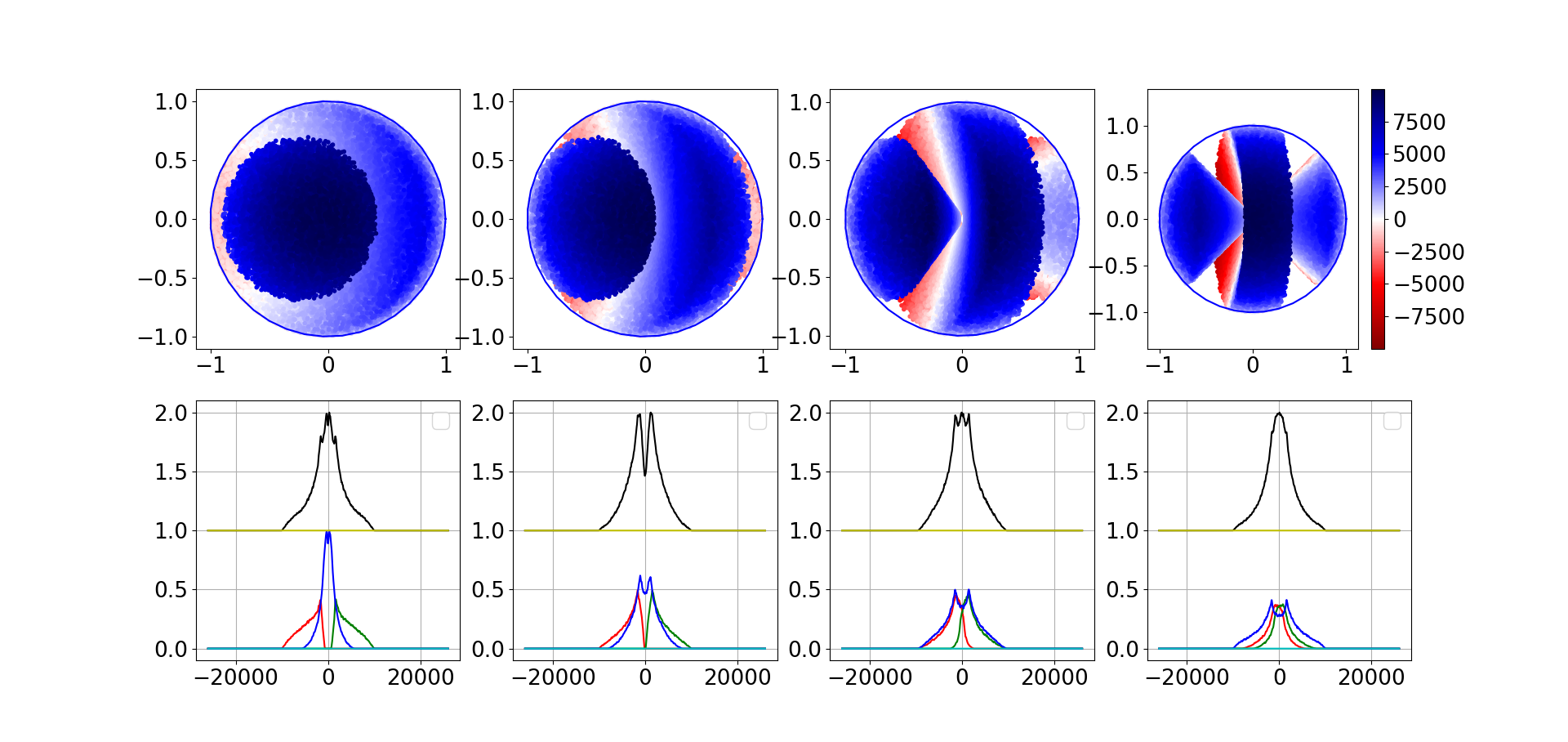}
    \caption{In the first row, the  layouts of the wind model BLR are represented for different inclinations: $i$~=~20°, 40°, 60° and 80°.
    Their respective spectra,  with no NLR emission added, are shown in the second row. 
    The color of the dots representing the emitting clouds are scaled as shown on the right-hand side of the top panels.
    The total spectrum (black line, middle row) is the sum of three contributions from the torus (blue line), the front cone (red line) and the back cone (green line) shown in the bottom row. }
    \label{fig:dispo_and_spectra}
\end{figure*}

\begin{figure}
\centering
\includegraphics[trim = 2.cm 0cm 2.5cm 1.5cm, clip,width=0.5\textwidth]{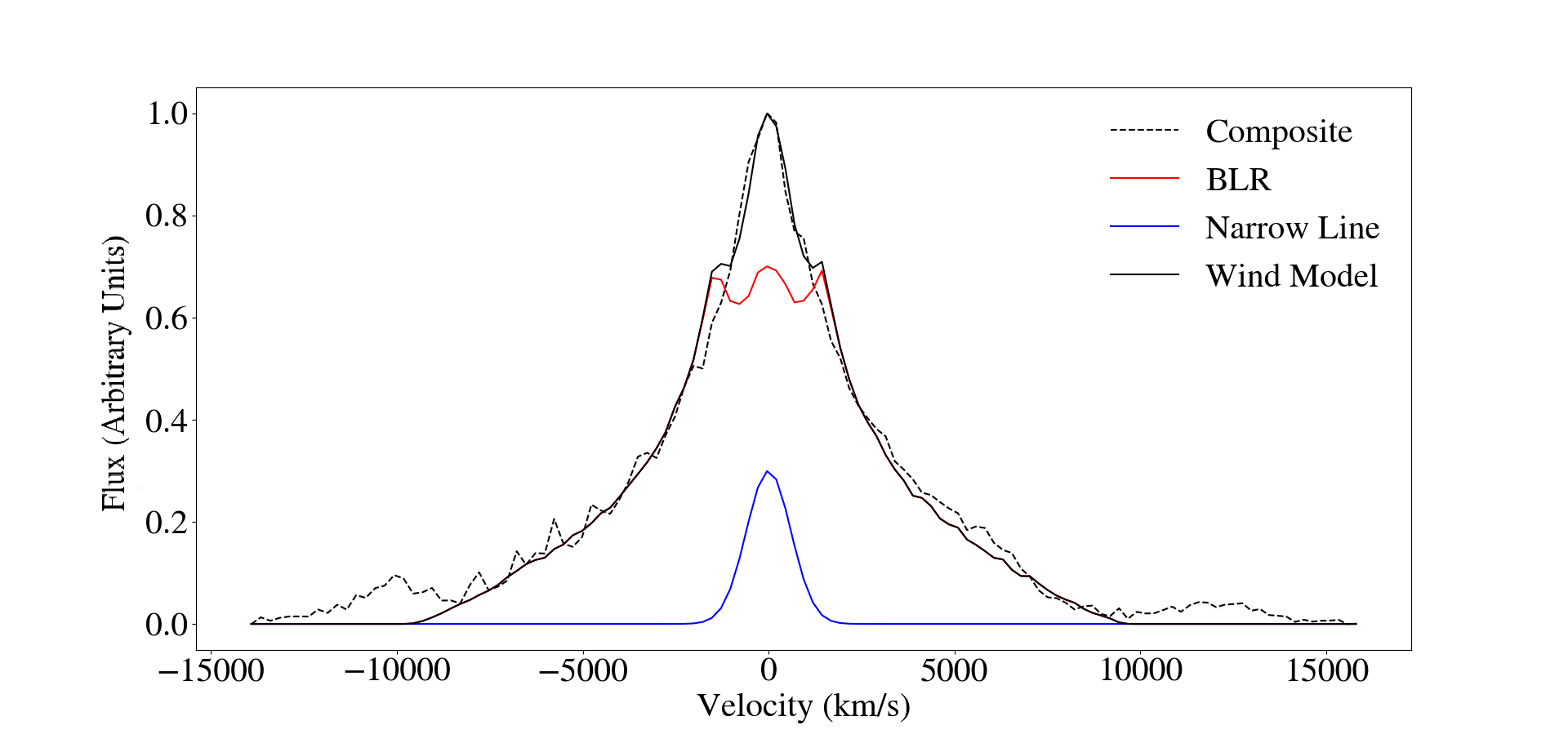}
\caption{Comparison of the composite spectrum \citep{composite} after subtraction of the quasar continuum and the N~{\sc v} emission (black dashed line) with a spectrum obtained with the wind model (see Text). The model (black solid line) is the sum of the BLR and NLR contributions (red and blue solid lines respectively). } 
\label{fit_composite_c}
\end{figure}

The second model is a combination of two models described in \citet{Braidant_2017}. We associate two winds, one equatorial and one polar (see Fig.~\ref{schema}). The velocity and density of the point-like clouds within the wind are described below.

As shown in blue on Fig.~\ref{schema}, the equatorial wind is a torus with an angle $a_{\rm torus}=15$° as in \citet{Braidant_2017}. In red, the polar wind is composed of two opposite cones with an opening angle $a_{\rm cone}=45$°.
The later value is a compromise. Indeed, a smaller value would result
in a gap on top of the emission line, when a larger value would mingle the contributions of the two winds implying a result close to the spherical distribution.
The cone and the torus are coaxial. The outer limit of both winds are $r_{\rm max}$~=~1 as for the spherical model but the inner radius is fixed at $r_{\rm min}=0.1$. The spectrum is much less sensitive to the later parameter as compared to the spherical model.

The point-like clouds in the wind flows have a radial velocity relative to the centre given by the following law: 
\begin{equation}
    v(r)= v_{\rm max}\, {\rm ln} \left( 1 + \frac{r}{r_{\rm max}} (e^1-1)\right)
    \label{eq: v cone}
\end{equation}
$v_{\rm max}$ is the maximal velocity of these clouds, reached at the maximal considered distance $r_{\rm max}=1$. 
Since the AGN winds are radiative pressure driven, the clouds are accelerating outwards.
The acceleration should decrease with the distance. This is why we chose a logarithmic velocity law for which the velocity is null at $r=0$ and {\rm which reproduces better the shape of the quasar Lyman-$\alpha$ emission}.
The \textit{ad hoc} factor $(e^1-1)$ was added only to fulfill the condition $v(r=r_{\rm max})=v_{\rm max}$.

The emitting clouds are not homogeneously spread inside the winds. The density of clouds is obtained by imposing the flux of clouds crossing the boundary of the layers  to be conserved  through the wind flow.
We use 750 layers and the first of them has 750 clouds in it, for a total of 350,252 emitting clouds inside the BLR. The number of layers and the number of clouds inside the first one are chosen such as the resulting spectrum is smooth and the wind model has a similar total number of clouds as the spherical one.

We then construct the observed spectrum by defining the observer position relative to the model axis. 
In this model, it must be noted that we only need one angle which is the inclination, $i$, of the axis relative to the line of sight to the observer. Indeed, by axial symmetry, all other positions will be recovered by a simple rotation.

Due to its peculiar geometry, the spectrum produced by the wind model varies as a function of the inclination, $i$, of the BLR. Fig.~\ref{fig:dispo_and_spectra} shows the spatial layout of the BLR and its corresponding spectrum for four values of the inclination : $i$~=~20°, 40°, 60° and 80°. In the same way as for the spherical model, the color of the dots indicates their velocity relative to the observer. However, due to the representation, it should be reminded that 
when projected in the same region of the sky, the blueshifted dots are hiding the redshifted ones, and thus for instance a DLA cloud located in the centre of a BLR with $i$~=~20° will not only obscure the blueshifted contribution but also the redshifted one not represented here. 
We have not added a NLR emission here to have a better insight on the contribution of each part of the BLR. 

At low inclination angle the absolute projected velocity of the clouds in the torus are small and accordingly the torus contribution is a narrow component centred at zero velocity whereas the contributions by the cones are spread at higher velocities and are well separated. When the inclination increases, the contribution by the torus is more spread over the velocities and is mixed with the contributions by the cones.

As an example, we show 
in Fig.~\ref{fit_composite_c} that we can reproduce the composite spectrum with typical parameters: $i=60$°, $v_{\rm max}=$9500~km~s$^{-1}$, the width of the narrow line is 500~km~s$^{-1}$, and the BLR/NLR strength ratio is 2.33.

\begin{figure}
\centering
    \begin{subfigure}[b]{0.5\textwidth}
    \includegraphics[trim = 0.cm 0cm 5.cm .cm, clip,width=\textwidth]{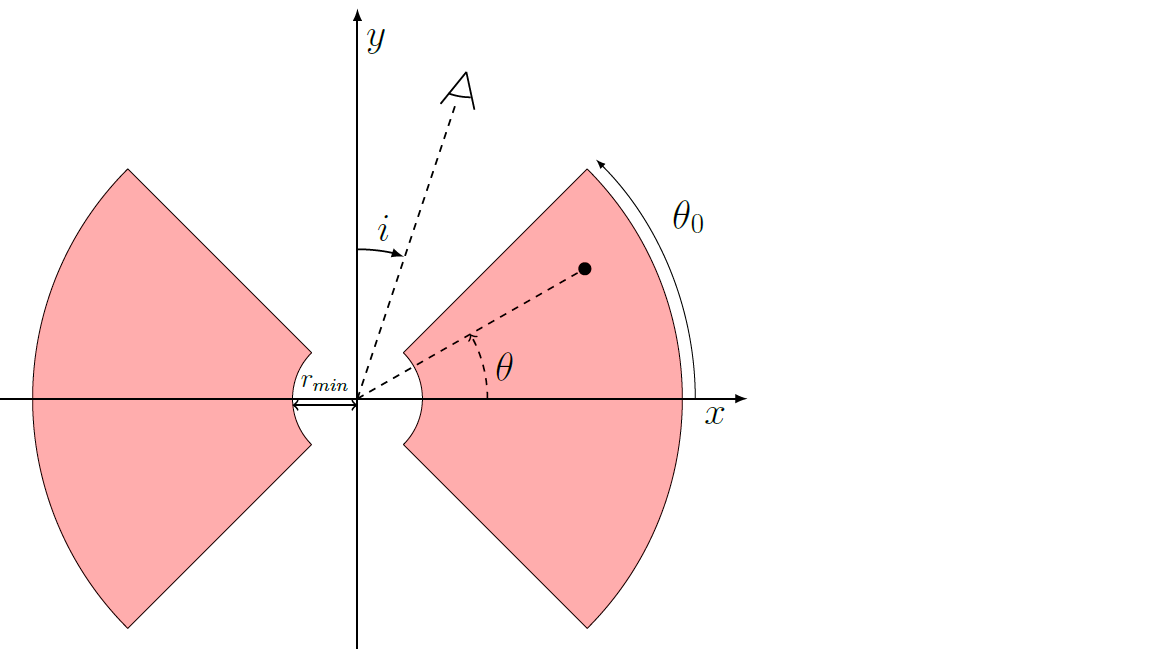}
    \caption{}
    \label{fig:schema_kepler}
    \end{subfigure}
    
    \begin{subfigure}[b]{0.5\textwidth}
    \includegraphics[trim = 4.5cm 1cm 3.cm 2.5cm, clip,width=\textwidth]{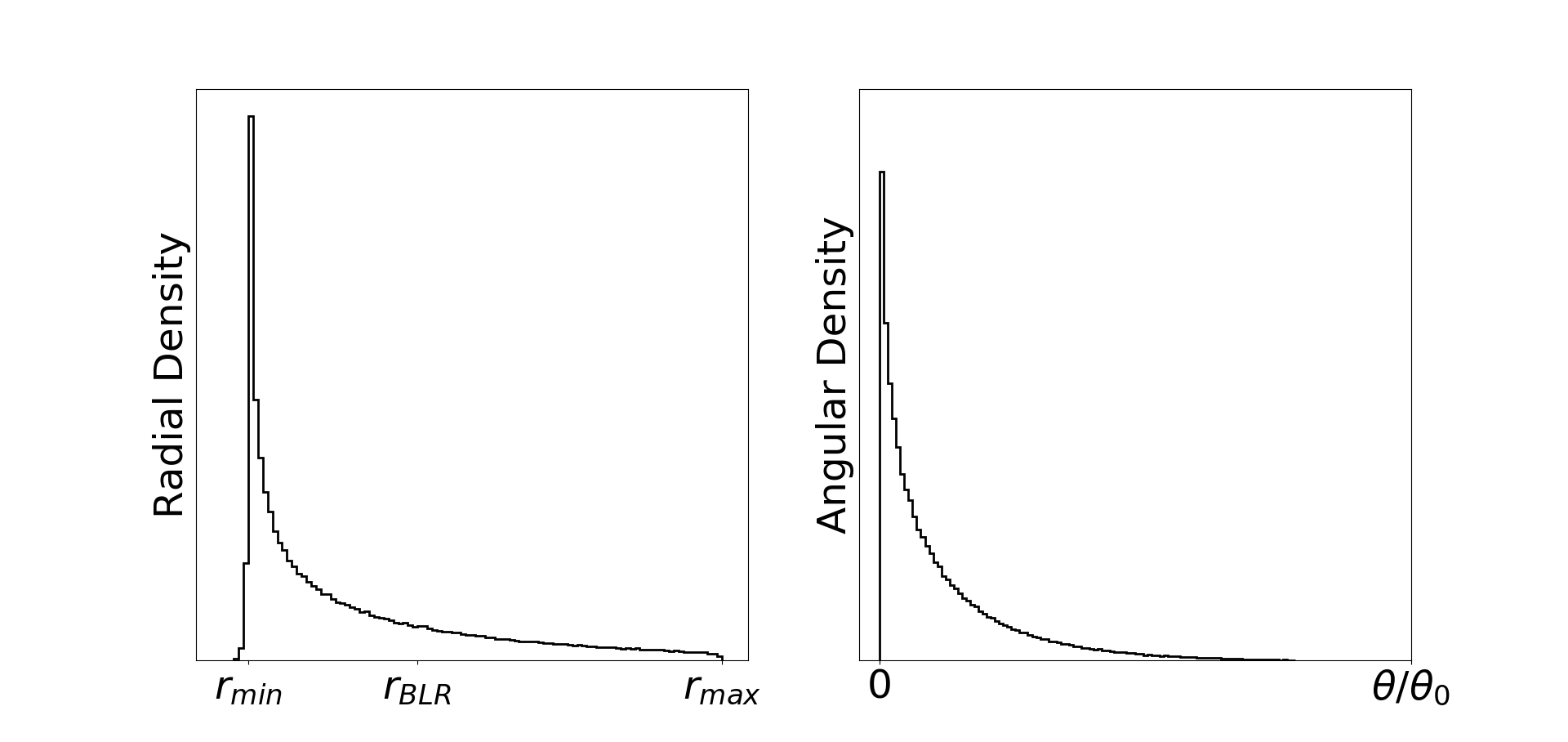}
    \caption{}
    \label{fig:densities_kepler}
    \end{subfigure}

\caption{\textit{Top panel}: Sketch of the cut view of the Keplerian disk model. The opening angle of the disk is $\theta_{0}=45$°. 
\textit{Bottom panels}: Radial (\textit{left}) and angular (\textit{right}) cloud density probabilities.} 
\label{fig:schema_kepler}
\end{figure}

\subsection{Keplerian disk model}

We use a simplified version of the model described in \cite{2014MNRAS.445.3055P} and favored by interferometric observations \citep{2018Natur.563..657G}. The model consists of a thick disk with half-opening angle $\theta_0$~=~45° in which point-like emitting clouds are moving along circular orbits around the black hole (see Fig.~\ref{fig:schema_kepler}). 
The emitting clouds are assigned a distance to the center with the following distribution.

\begin{equation}
    r = r_{\rm min} + g\left(\frac{1}{\beta^2}\right)\times (1-F) 
    \beta^2 \, r_{\rm BLR}
    \label{eq: r_distribution_kepler}
\end{equation}

where $g\left(\frac{1}{\beta^2}\right)$ is a gamma distribution with a shape parameter $\beta=1.4$ and $F = r_{\rm min} / r_{\rm BLR}$ with $r_{\rm min}=0.1$ and where $r_{\rm BLR}$ is the mean radius of the BLR. We use the dimensions of the model by \cite{2018Natur.563..657G} so that $r_{\rm BLR} = 0.42$. As for other models, the maximum radius of the BLR is $r_{\rm max}~=~1.$

The azimuth $\theta$ of the clouds follows an exponential probability distribution of scale height $\theta_0$/4 as presented on Fig.~\ref{fig:densities_kepler}.

The velocity of the clouds is given by 
\begin{equation}
    v(r) = v_0 \times \sqrt{\frac{r_{\rm min}}{r}}
    \label{eq: velocity_kepler}
\end{equation}
$v_0$  being a free parameter. The direction of the velocity is perpendicular to the radius of the orbit. The total number of clouds in the BLR is 350,000 such as it matches with the previous two other models.

\begin{figure}
    \centering
    \includegraphics[trim = 3.cm 0cm 3.5cm 2.5cm, clip,width=0.5\textwidth]{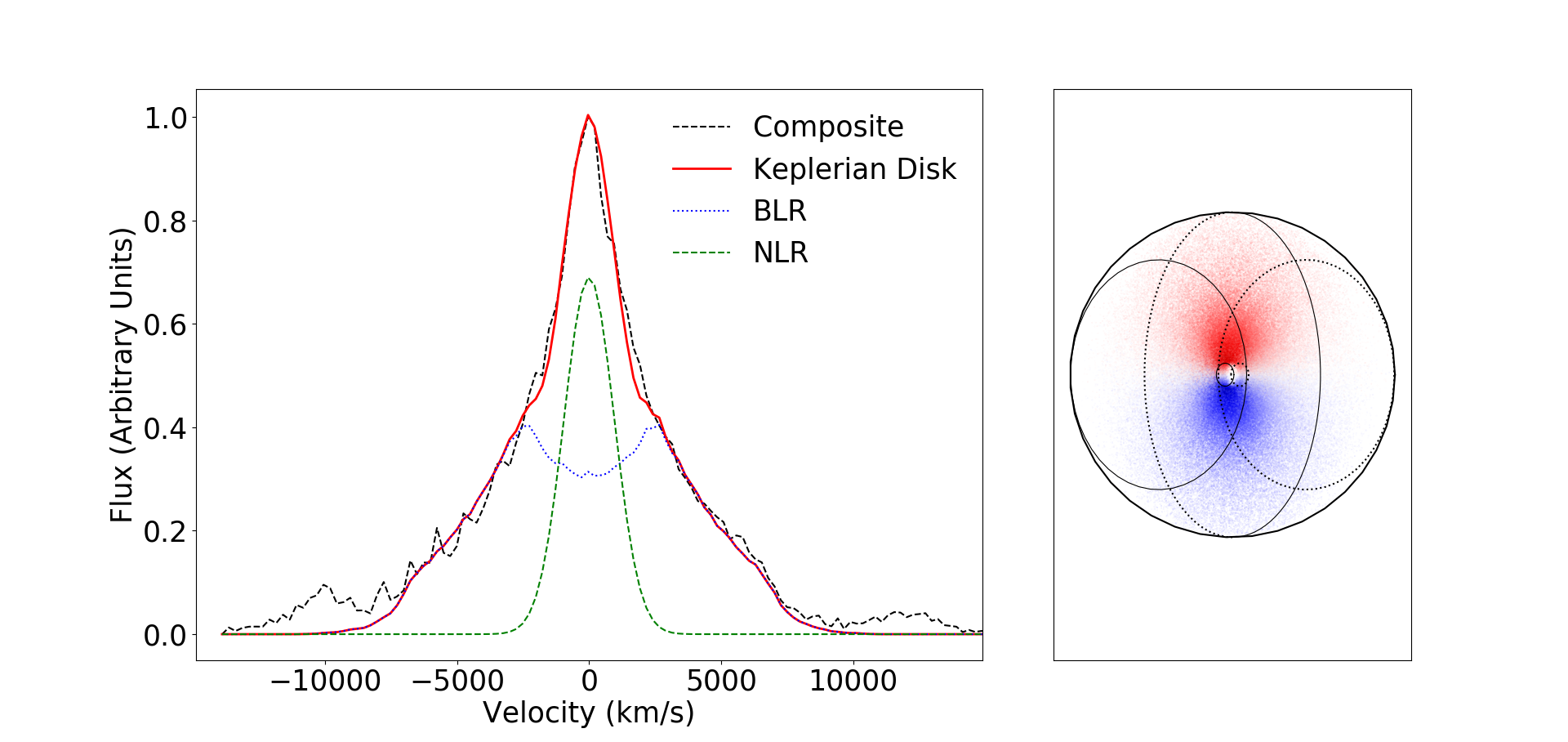}
    \caption{ {\sl Left} : Fit of the quasar composite spectrum 
    by the Keplerian disk model. The quasar continuum and the N~{\sc v} emission have been subtracted beforehand.
    Total, BLR and NLR spectra are represented by red, blue and green curves, respectively. {\sl Right}: The relative projected velocities of the emitting clouds are     represented by dots with colors indicating their direction. The inclination of the disk is $i=40$°. To ease the visualisation, the edges of the disk are represented by full lines when they are on the side of the observer and dotted lines when they are hidden on the opposite side.}
    \label{fig:kepler_fit}
\end{figure}

As it can be seen on the right-hand side panel of Fig.~\ref{fig:kepler_fit}, blueshifted clouds are located on one side of the plane when redshifted clouds are located on the other side.

The left-hand side panel of Fig.~\ref{fig:kepler_fit} shows the fit of the quasar template with this model. It can be seen on the figure that the BLR spectrum shows two peaks widely separated implying that the needed NLR emission has a broader width, $FWHM~=~900$~km~s$^{-1}$ in this case.
The main parameter of the model is the inclination, $i$, between the disk axis and the line of sight to the observer.

\section{Ghostly DLAs}

A \textit{ghostly}-DLA is the result of the presence of a small absorbing cloud in front of the BLR. The cloud is small enough so that part of the BLR is not covered. 
One very important observational fact to bear in mind is that the cloud must cover the central source of continuum. Indeed, \textit{ghostly}-DLAs are identified by the presence of strong metal absorption lines some of them being redshifted in spectral regions devoid of any emission line.

After placing the DLA-cloud in front of the BLR, we define which emitting clouds are covered and which are not. We derive the total emission of the covered region and apply  to the resulting spectrum the absorption by the amount of neutral hydrogen in the cloud. We then add  to the absorbed spectrum the contribution of the uncovered part of the BLR.

For simplicity, we consider a cylindrical absorbing cloud of radius $r_{\rm cloud}$ and constant column density. 
The resulting spectrum depends on several characteristics of the absorbing cloud: its column density (which can be estimated from the Lyman series absorptions when these lines are seen in the quasar spectrum), its position, its size; but it depends also on the inclination of the BLR with respect to the observer in the case of the wind and disk models.

In the following, we illustrate the impact of an absorbing cloud on the modelled quasar spectra. 
We impose the BLR models to reproduce the template quasar emission and therefore fix parameters so that models do so (see previous Section). 
\begin{figure*}
\centering
    \includegraphics[
    trim = 3cm 0.cm 4.5cm 2.5cm,
    clip,width=0.90\textwidth]{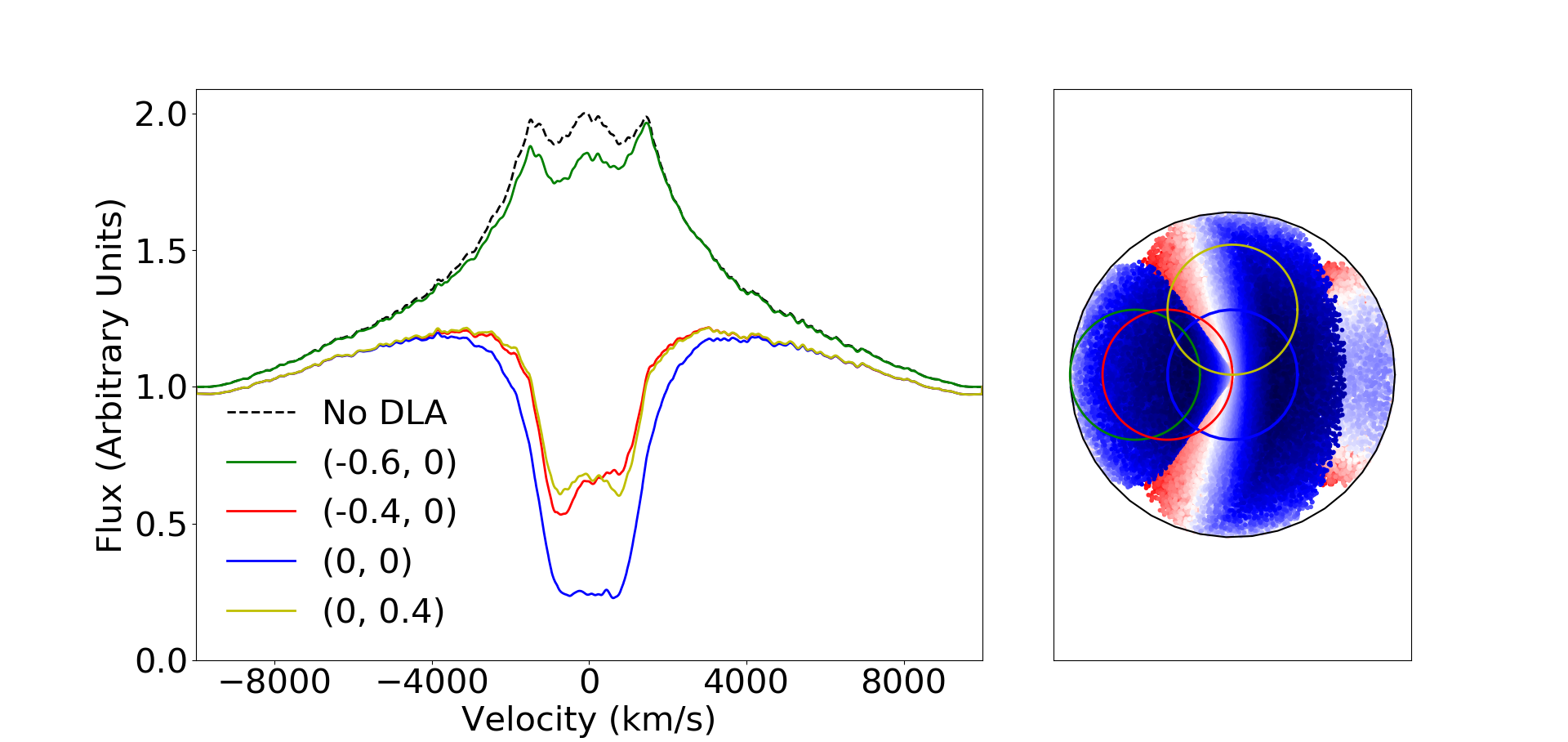}
    \caption{Left panel shows spectra built with the wind model (with $i$~=~60°) and an  absorbing cloud located at different  positions as indicated in the right panel. The absorbing cloud has a column density log~$N$(H~{\sc i})(cm$^{-2}$)~=~21 and a radius $r_{\rm cloud}$~=~0.4. 
    The quasar spectrum without any absorption and with no NLR emission is represented with a dotted black line.
    }
    \label{fig:dif_position}
\end{figure*}

For this exercise, we assume that the quasar continuum has the same density flux as the top of the Lyman-$\alpha$ emission line which is typical of bright quasars at these redshifts and that the column density of the absorbing cloud is log~$N$(H~{\sc i})(cm$^{-2}$)~=~21. In addition we intentionally minimize the flux from the NLR component to illustrate better the consequences of partial coverage of the BLR. In the two following subsections, we do not add any NLR emission.

\begin{figure*}
    \includegraphics[
    trim = 3cm 0.cm 4.5cm 2.cm,
    clip,width=0.90\textwidth]{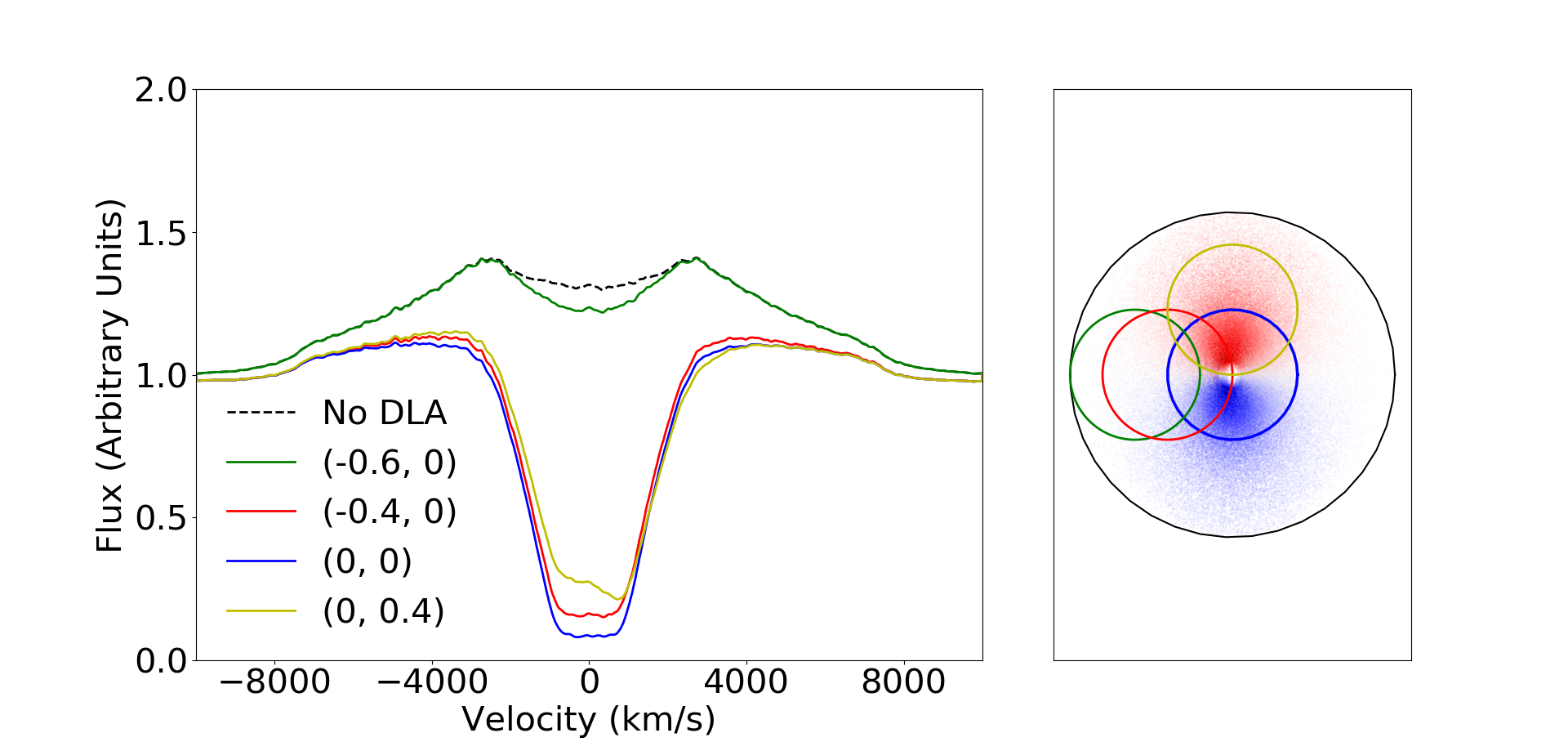}
    \caption{ Left panel shows spectra built with the disk model (with $i$~=~40°) and an absorbing cloud located at different  positions as indicated in the right panel. The absorbing cloud has a column density log~$N$(H~{\sc i})(cm$^{-2}$)~=~21 and a radius $r_{\rm cloud}$~=~0.4. 
    The quasar spectrum without any absorption and with no NLR emission is represented with a dotted black line.
    }
    \label{fig:dif_position2}
\end{figure*}

\subsection{Position of the cloud}

Given the symmetries of the models, the position of the absorbing cloud has  more impact on the resulting spectrum for the wind  and disk models.

In the left panel of Fig.~\ref{fig:dif_position}, we present spectra obtained with the wind model assuming an absorbing cloud located at the different positions indicated in the right panel. The inclination of the cone is 60° and the cloud has a radius $r_{\rm cloud}$~=~0.4. The unabsorbed quasar spectrum is shown as the dotted black line and the green line represents the spectrum of the quasar with the cloud centred at $(-0.6,0)$, thus not covering the source of continuum located in the centre. The other positions of the cloud are at $(-0.4,0)$ (red solid line), $(0,0)$ (blue solid line) and $(0,0.4)$ (yellow solid line). Note that, by symmetry, the spectrum will be the same if the cloud is centred at $(0,-0.4)$ or at $(0,0.4)$.  For the same reason, when the cloud is at $(0.4,0)$ the spectrum will be the mirror version (relative to the zero velocity) of the spectrum when the cloud is at $(-0.4,0)$. 
Indeed, the covered BLR emitting clouds moving toward the observer in one case are moving away in the other case. One can notice that the absorption is more important when the cloud is centred at  $(0,0)$. This is due to the higher density of emitting clouds at small distances from the centre. Only a small fraction of these numerous low-velocity emitting BLR clouds are covered by the absorbing cloud when located far from the centre. One can also notice the asymmetry in the  $(-0.4,0)$ spectrum due to the majority of emitting clouds with negative velocity covered whereas the  $(0,0.4)$ spectrum is symmetric due to the same number of emitting clouds with negative and positive velocities covered. One can argue nonetheless that the difference between the two cases is rather small but in other situations the difference could be more significant. 

In the left panel of Fig.~\ref{fig:dif_position2} we present the spectra obtained with the disk model assuming an absorbing cloud located at the different positions indicated in the right panel which are identical to those used for the wind model. The spectra look similar to that of the wind model but with a symmetry relative to the y-axis instead of a symmetry relative to the x-axis. It is apparent however that, because of the large opening angle of the disk, the red and blue peaks in the corresponding quasar spectrum are less absorbed resulting in the wings of the absorption trough to be steeper.
The resulting two emission peaks on both sides of the absorption are more distant compared to the wind model. This implies that the NLR emission, needed to fill the residual absorption, will have to be broader for these models than for the wind model. This could imply that for a fixed radius the absorbing cloud should be closer to the AGN in order to avoid absorption of the central part of the NLR where velocities are expected to be larger.

\begin{figure}
    \centering
    \includegraphics[trim = 2.5cm 0cm 3cm 1cm, clip,width=0.5\textwidth]{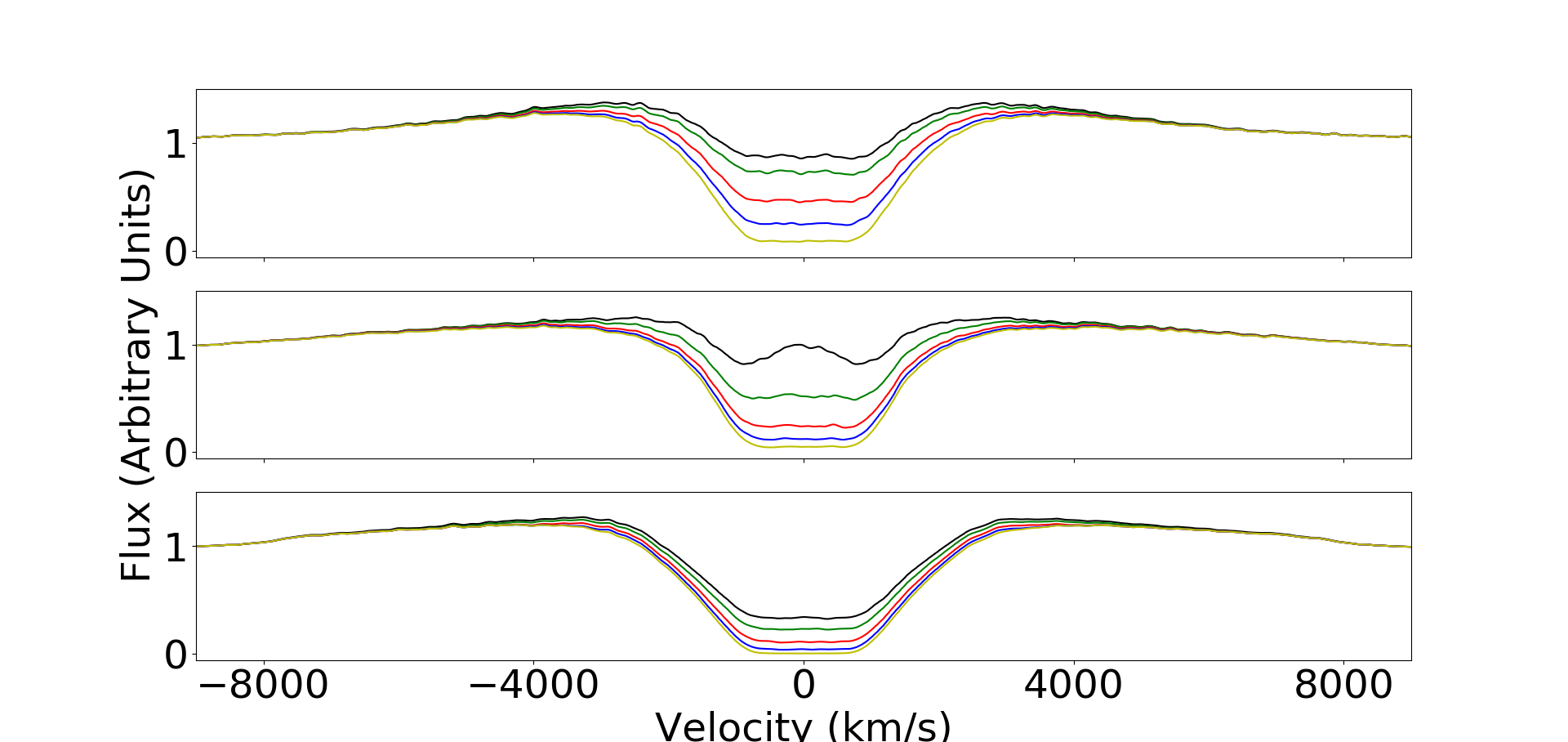}
    \caption{Spectra obtained with the spherical model (upper panel), the wind model (middle panel) and the Keplerian disk model (lower panel) with an absorption cloud located at (0,0) and with a radius of 0.1 (black line), 0.2 (green line), 0.4 (red line), 0.6 (blue line) and 0.8 (yellow line).}
    \label{fig:dif_rad}
\end{figure}

\subsection{Size of the cloud}

Fig.~\ref{fig:dif_rad} shows how the spectrum evolves with the size of an absorbing cloud centred at (0,0) for the three models, the spherical model (upper panel), the wind model (middle panel)  and the disk model (bottom panel). 

The continuum and the BLR flux levels have been fixed to 1 and the absorbing cloud column density is still log~$N$(H~{\sc i})~=~21.0. For the wind and disk models, the inclination of the model axis is 60° and 40°, respectively. As one could expect, the absorption is getting more prominent when the size of the absorbing cloud increases.
One can see that it is easy to reproduce a \textit{ghostly}-DLA for the spherical and wind models without tuning the parameters. It is possible to hide the absorption even more by decreasing the ratio between the continuum and the BLR fluxes.
This is obtained without adding a NLR emission which is not absorbed and would fill in part if not all of the residual absorption.
To obtain a \textit{ghostly}-DLA with the disk model is more difficult and a stronger NLR is needed.
 
\subsection{Examples}

\begin{figure}
    \centering
    \includegraphics[trim = 4cm 3cm 3cm 1cm, clip,width=0.5\textwidth]{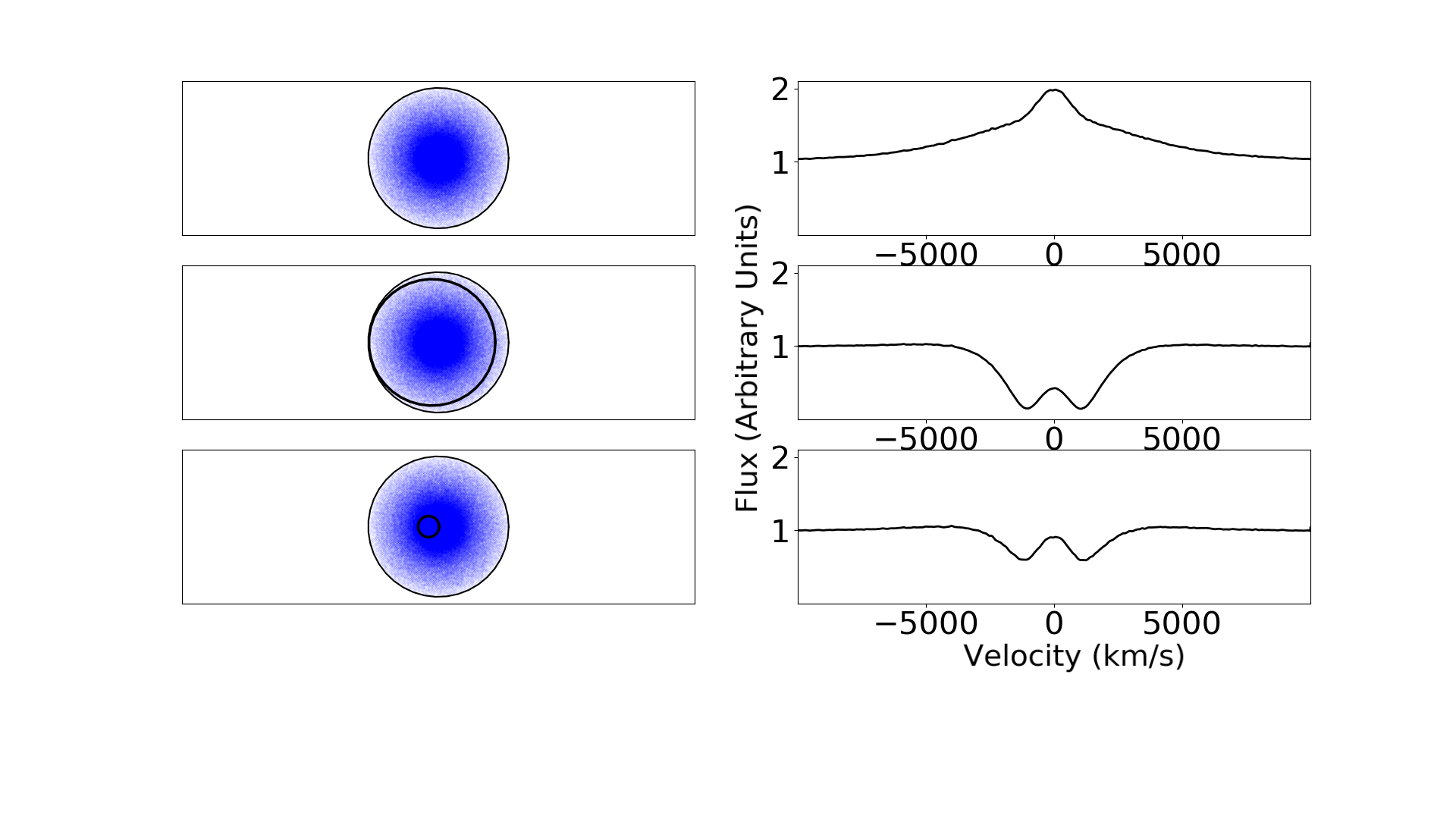}
    \caption{Examples of spherical model spectra (right-hand side column) and their corresponding spatial structure (left-hand side column). {\sl Top to bottom}: Regular QSO, eclipsing DLA QSO, \textit{ghostly}-DLA QSO.}
    \label{fig:examples_s}
\end{figure}

\begin{figure}
    \centering
    \includegraphics[trim = 4cm 3cm 3cm 1cm, clip,width=0.5\textwidth]{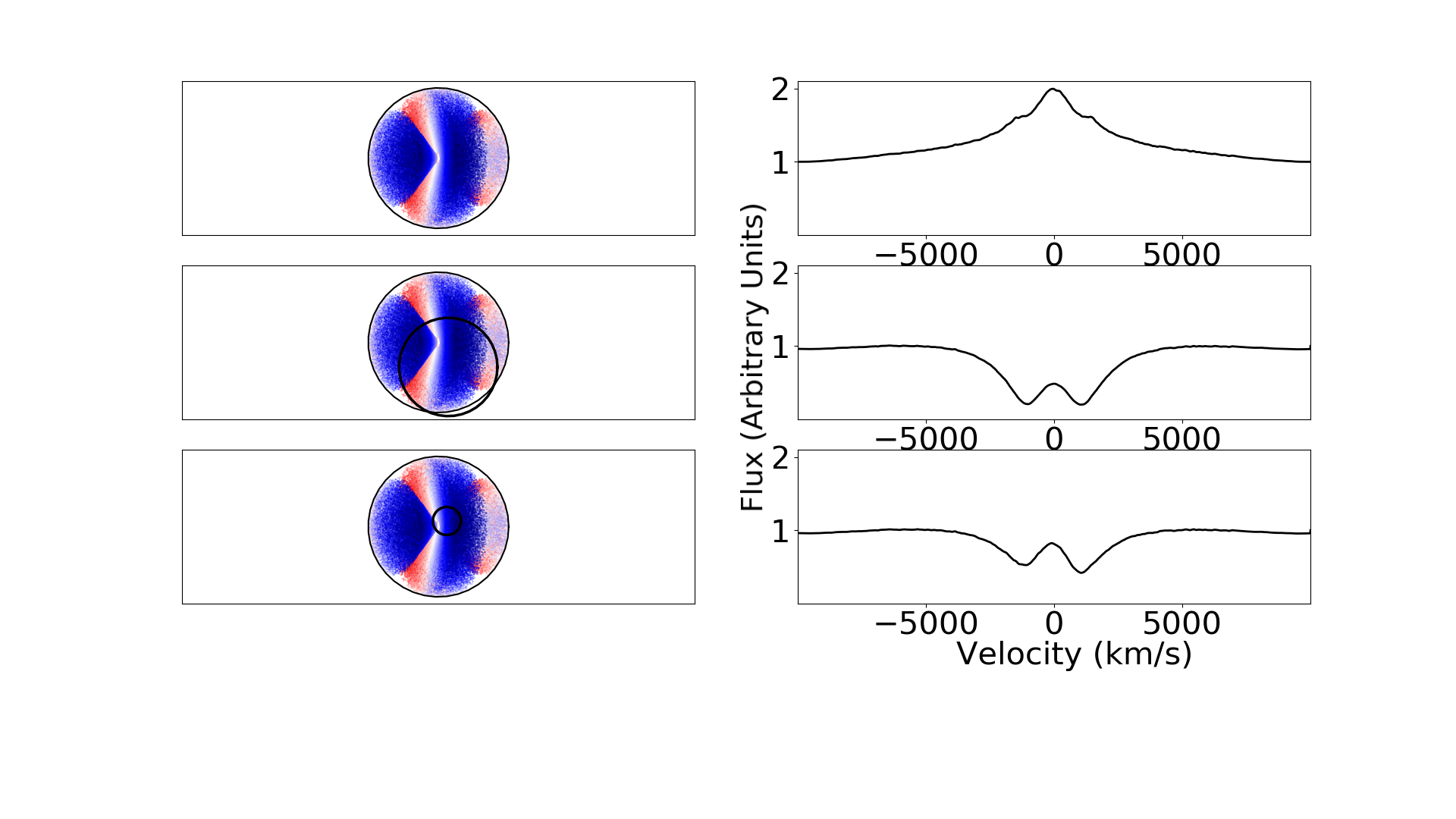}
    \caption{Examples of wind model spectra (right-hand side column) and their corresponding spatial structure (left-hand side column). {\sl Top to bottom}: Regular QSO, eclipsing DLA QSO, \textit{ghostly}-DLA QSO}
    \label{fig:examples_w}
\end{figure}

\begin{figure}
    \centering
    \includegraphics[trim = 4cm 3cm 3cm 1cm, clip,width=0.5\textwidth]{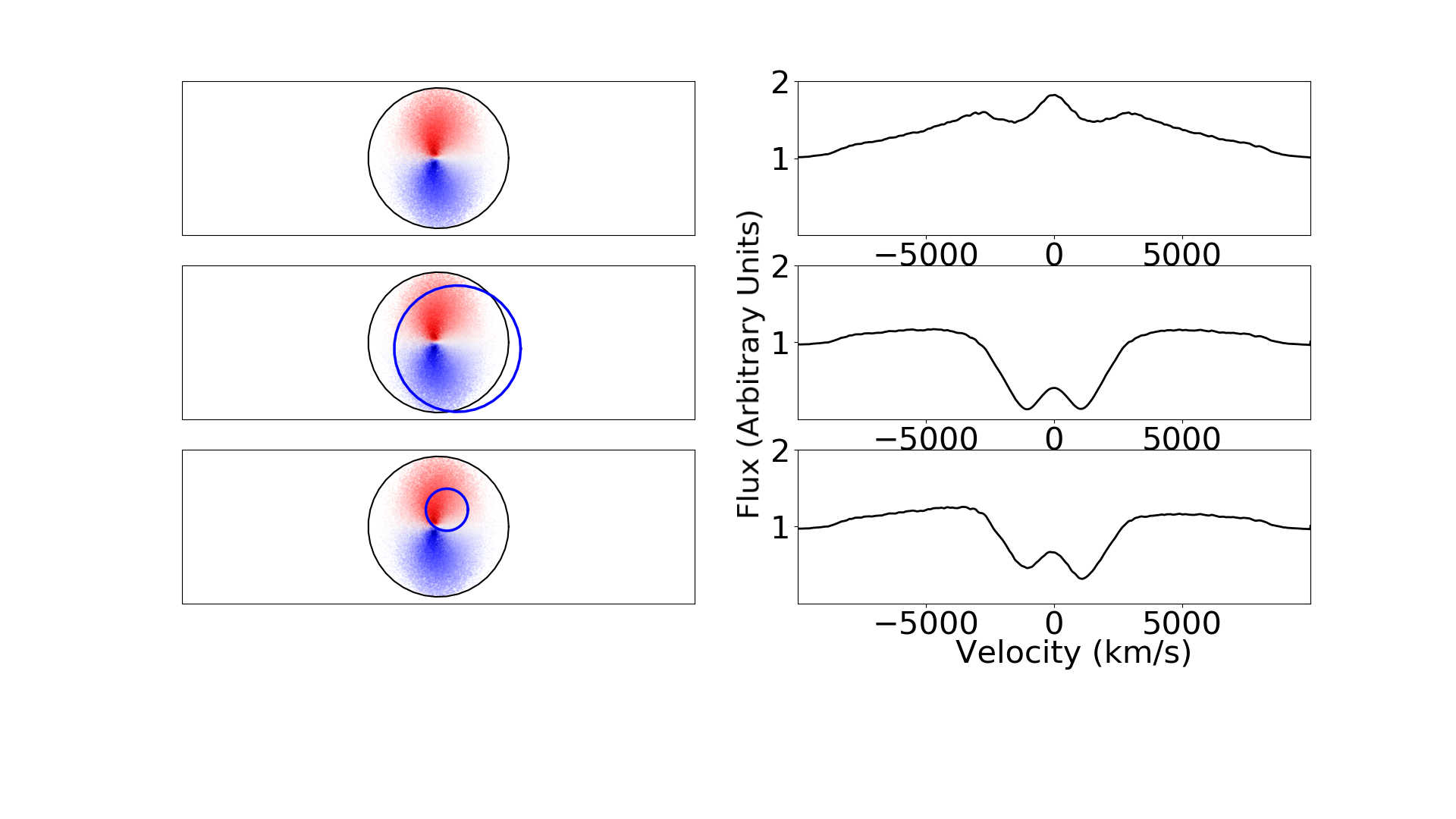}
    \caption{Examples of Keplerian disk model spectra (right-hand side column) and their corresponding spatial structure (left-hand side column). {\sl Top to bottom}: Regular QSO, eclipsing DLA QSO, \textit{ghostly}-DLA QSO}
    \label{fig:examples_k}
\end{figure}

With the models we can tune the parameters to obtain spectra of different types of quasar Lyman-$\alpha$ emission lines. 
As said before and derived from observations, we impose the cloud to cover the quasar source of continuum. We also add a weak NLR emission.
In Fig.~\ref{fig:examples_s}, Fig.~\ref{fig:examples_w} and  Fig.~\ref{fig:examples_k} for the spherical, wind and disk models respectively, we show the spectrum of a quasar with no absorption (right-hand side upper panel), and the same with an absorbing cloud in front (two other panels). The corresponding spatial structure is shown in the left-hand side panels. It can be seen that the spectra in the middle panels correspond to an eclipsing DLA, where the absorbing cloud behaves as a coronagraph and only a weak narrow Lyman-$\alpha$ emission is seen in the bottom of the trough \citep{2013A&A...558A.111F}.
The spectra in the bottom panels correspond to \textit{ghostly}-DLAs.
To obtain an eclipsing DLA-QSO in the case of the spherical and disk models, the radius of the absorbing cloud must be large enough to cover a significant portion of the BLR. 
Whereas the cloud can be smaller in the case of the wind model geometry.

On the other hand, to obtain a \textit{ghostly}-DLA, the absorbing cloud must be rather small so that the non-covered emission fills up at least part of the absorption. A high BLR flux relative to the continuum flux also helps to obtain such \textit{ghostly}-QSOs. More importantly a strong NLR emission can fill in the trough as soon as the width of the DLA absorption trough matches the width of the NLR emission.

The evolution of the models as a function of the different parameters is discussed in more details in the next section.


\section{Investigation of the models}\label{section: Investigation}

Our main objective is to extract any information on the BLR structure and on the characteristics of the absorbing cloud from observations of \textit{ghostly}-DLA QSOs by comparing the quasar spectra with the outputs of our models. 
Before performing direct comparison, we would like to gain insight on which parameters can be constrained best. To do so, we will construct realistic mock spectra and fit them back with our models.

\subsection{Mock spectra}

Mock spectra are built from the models described earlier. We impose parameters so that the modelled emission spectrum fits the quasar composite spectrum. 
We then choose the parameters of the absorbing cloud: its size, position and column density, in such a way that the corresponding spectrum belongs to the \textit{ghostly}-DLA category. An important characteristic of these spectra is that the source of the quasar continuum located at the center of the models must be covered by the absorbing cloud. Indeed, strong metal lines are observed associated with \textit{ghostly}-DLAs some of them redshifted in wavelength ranges devoid of emission lines.

Noise is added to the modelled spectrum with a given signal-to-noise ratio (SNR) and the spectrum is rebinned to a spectral resolution $R$.
For each model, we will consider spectra with $R$~=~2,500 and SNR~=~10 on the one hand and $R$~=~5,000 and SNR~=~50 on the other. 
The first case (LR for low resolution) corresponds roughly to the characteristics of good SDSS spectra. The second case (HR for high resolution) investigates 
what could be done with better data that could be obtained with e.g. XSHOOTER on the VLT.
Given the width of the lines, higher spectral resolution is not needed.

The chosen parameters for the models are presented in the first row of Table.~\ref{tab:from_spherical},  Table.~\ref{tab:from_conetore}  and  Table.~\ref{tab:from_KN} for the spherical, wind and disk models respectively.

\subsection{Fit of mock spectra}

Once a mock spectrum has been generated, it is fitted with the three models in order to estimate the degeneracy between models and to evaluate our capability to recover some of the true parameters.
For this, we compute the reduced $\chi^2$ between the mock spectrum and the models  built with every possible parameter combination. The reduced $\chi^2$ is computed as follows:
\begin{equation}
    \chi^2=\frac{1}{n-m} \sum_i \frac{(O_{\rm i}-M_{\rm i})^2}{\sigma_i^2}
    \label{chi2}
\end{equation}
where $n$ is the number of pixels, $m$ the number of fitted parameters, $O_{\rm i}$ the observation, $M_{\rm i}$ the model and $\sigma_i^2$ the variance of the data. 
As the far wings of the emission are not well reproduced by our simplified models,the $\chi^2$ computation is performed taking into account only the pixels between $-7000$~km~s$^{-1}$ and $+7000$~km~s$^{-1}$.

We vary the parameters as follows:
\begin{itemize}
    \item The radius of the absorbing cloud: every tenth of a distance unit from 0.1 to 0.9.
    \item The x coordinate of the quasar: every tenth of the radius of the cloud, going from 0 to $r_{\rm cloud}$ for the spherical  and disk models and from $-r_{\rm cloud}$ to $r_{\rm cloud}$ for the wind model.
    \item For the wind and disk models, the y coordinate: every tenth of the radius of the cloud such as $\sqrt{x^2+y^2}<r_{\rm cloud}$.  For the wind model, $y \ge 0$ whereas it can be positive or negative for the disk model.
    \item For the wind and disk models, the inclination of the BLR: every ten degrees, from 0° to 90°.
    \item The column density: 12 values between log~$N$(H~{\sc i})~=~19.0 to 21.4.
    \item  The strengths of the NLR and BLR emissions are varied only slightly to optimize the fit.
\end{itemize}

Note that we fix the width of the NLR to $FWHM_{\rm NLR}=600$~km~s$^{-1}$ and the maximal velocity $v_{\rm max}=10,000$~km~s$^{-1}$ for the spherical and wind models. For the disk model, the width of the NLR component is also fixed but at $FWHM_{\rm NLR}=950$~km~s$^{-1}$.

We vary the free parameters and compare the mock spectra with hundred of thousands of models. In reality, note that the column density for some \textit{ghostly}-DLAs with high enough redshift could be inferred from the 
Lyman series. We however want to check if our method can recover the
correct column density in case these lines are not available in the observed wavelength window.

\subsection{Spherical model}

\begin{table*}
    \centering
    \begin{tabular}{c|c|c|c|c|c|c|c|c|c|c|c}
         \textbf{\#}&\textbf{Method}&\textbf{Quality} & \textbf{$N$(H~{\sc i})} & \textbf{Radius} & \textbf{x Coordinate}& \textbf{y Coordinate} &\textbf{Inclination}& \textbf{Strength NLR} & \textbf{Strength BLR}& \textbf{$\chi^2$}\\
    \hline
    \hline
    
    1&Spherical& N.A. & 20.6 & 0.3 & 0.0 & 0.0 & N.A. & 0.60 & 1.40 & N.A.  \\
    \hline\hline
    2&Spherical & LR & \textbf{20.6} & 0.4 & 0.24 & \textbf{0.0} & N.A. & \textbf{0.59} & \textbf{1.40 }& 1.01\\
    3&Spherical& HR  & \textbf{20.6} & 0.5 & 0.12 & \textbf{0.0} & N.A. & \textbf{0.66} & \textbf{1.39} & 1.13\\
    \hline
    4&Wind & LR & 21.0 & 0.2 & \textbf{0.0} & 0.18 & 50 & 0.29 &  1.89 & 1.03\\
    5&Wind & HR & 21.0 & 0.1 & \textbf{0.0} & \textbf{0.0} & 50 & 0.01 &  1.81& 2.11 \\  
    \hline
    6&Disk & LR & 20.3 & 0.2 & 0.04 & -0.02 & 30 & 0.88 &  0.90 & 1.06\\
    7&Disk & HR & 20.3 & 0.1 & 0.07 & -0.07 & 30 & 0.79 &  0.91& 3.52 \\  
    \hline
    \hline
    8&Wind & LR & \textbf{20.6*} & 0.9 & 0.09 & 0.81 & 50 & 0.86 &  1.58 & 1.11\\
    9&Wind & HR & \textbf{20.6*} & 0.8 & \textbf{0.0} & 0.72 & 50 & 0.93 &  1.57& 4.12 \\ 
    \hline
    10&Disk & LR & \textbf{20.6*} & 0.9 & 0.09 & 0.18 & 0 & 1.11 &  5.37 & 1.13\\
    11&Disk & HR & \textbf{20.6*} & 0.2 & 0.18 & -0.04 & 20 & 0.68 &  1.18& 4.72 \\ 
    \end{tabular}
    \caption{Results from the fit of a mock spectrum constructed with the spherical model in two versions LR and HR. The input parameters are indicated in the first row. We fit the low and high resolution spectra with the spherical, wind and disk models to try to recover the input parameters. The parameters of the best fits for the different models are presented in rows \#2 to \#7. 
    When the fit recovers the initial parameter within 10\%, the value is printed in boldface. The second part of the Table (rows \#8 to \#11) shows the same with the neutral hydrogen column density fixed at the correct value (20.6) as indicated by an asterisk.}
    \label{tab:from_spherical}
\end{table*}
Parameters from which we construct the mock spectrum of the spherical model are listed in the first row of Table~\ref{tab:from_spherical}. Since the BLR is a sphere, no inclination is needed but also, due to axial symmetry, the cloud is only moved along the x-axis, and the y coordinate is kept equal to $0$. 

We then fit the mock spectrum in its two versions, LR and HR, with the spherical, wind  and disk models. Results of the best fits are given in Table~\ref{tab:from_spherical} from row \#2 to row \#7.

The fits of the LR spectrum are equally good for all models due to the noise hiding the differences between the models. 
On the other hand, not surprisingly, the spherical model gives the best fit in HR. 
This is encouraging because this exercise shows that we may be able to distinguish between the three models providing good data with sufficiently high spectral resolution and SNR are available.

When the redshift of the system is high enough, the absorptions from the other Lyman series lines are seen in the quasar spectrum \citep{Fathivavsari_2017} and the neutral column density can be derived directly from these absorptions. We have therefore fixed the column density to the correct value and reproduce the exercise. The results are presented in Table~\ref{tab:from_spherical} from row \#8 to row \#11. We can notice higher $\chi^2$ values for the wind and disk models which makes the spherical model even more distinguishable from the other two models.

The ratio between BLR and NLR emissions is approximately retrieved. This is however not the case for the size of the DLA-cloud and its position even for the spherical model in HR. We will discuss further in the next section the constraints derived on the parameters.

\begin{figure}
\begin{subfigure}[b]{0.5\textwidth}
    \includegraphics[trim = 3.cm 0.cm 4.5cm 2.5cm, clip,width=\textwidth]{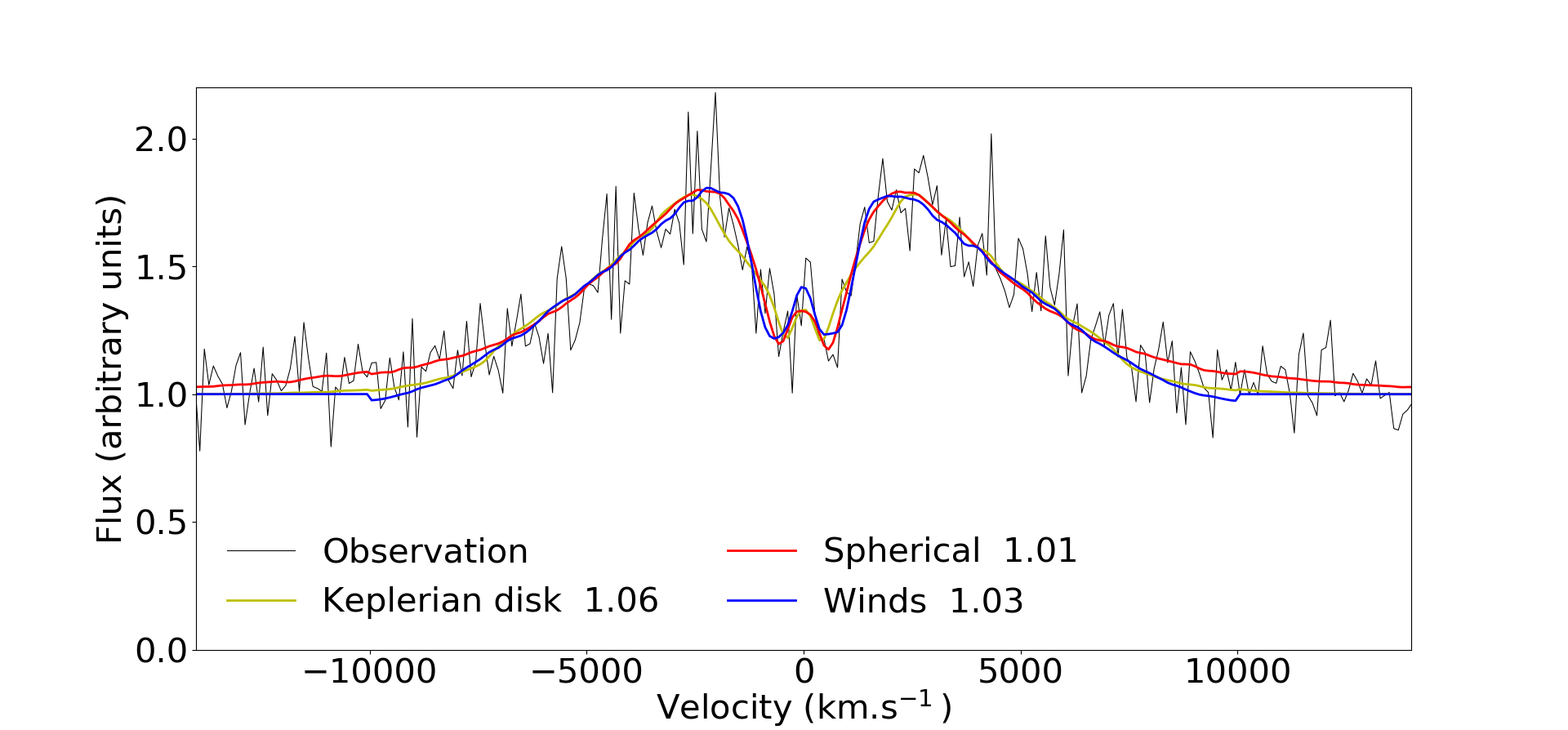}
    \caption{}
    \label{fig:compa_A}
\end{subfigure}

\begin{subfigure}[b]{0.5\textwidth}
    \includegraphics[trim = 3.cm 0.cm 4.5cm 2.5cm, clip,width=\textwidth]{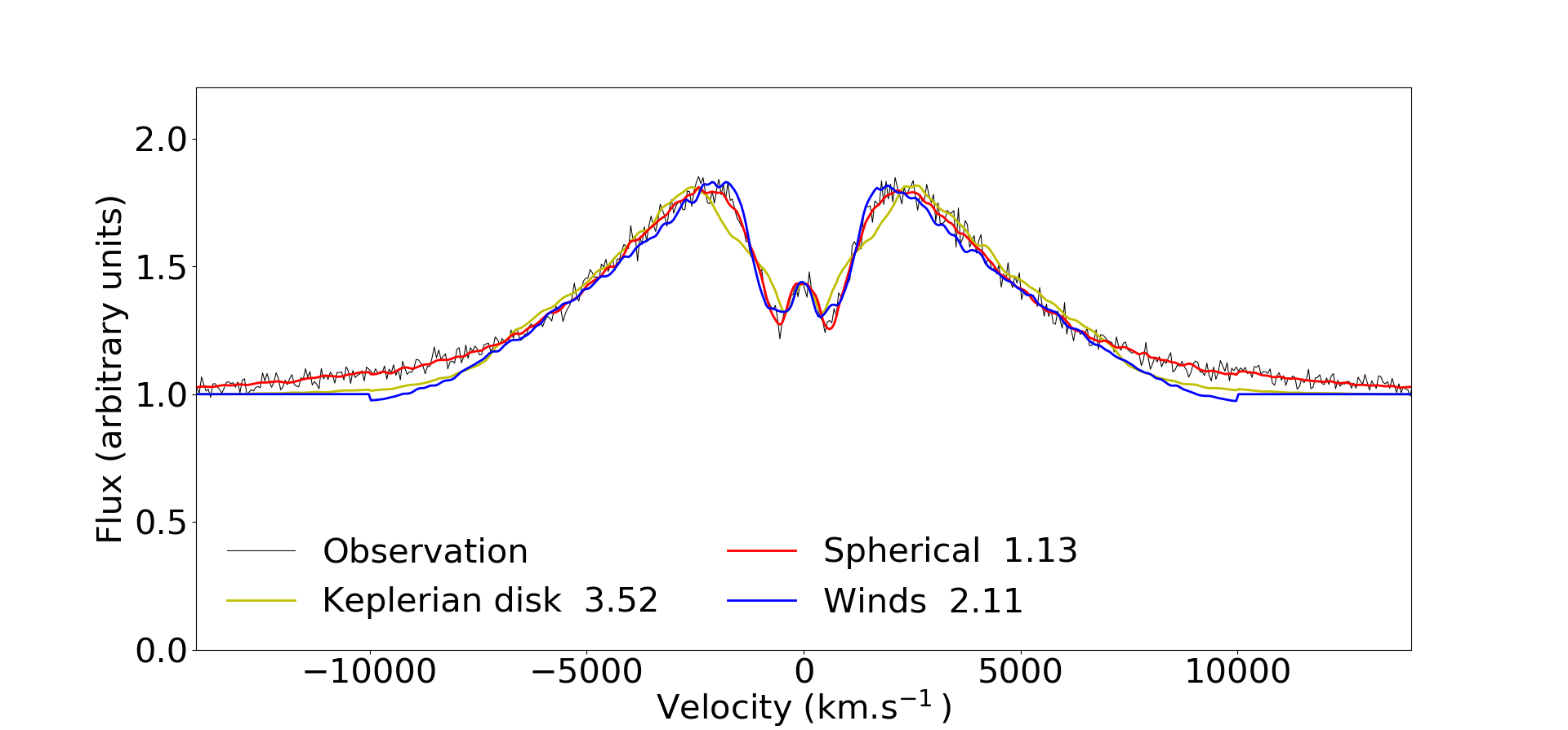}
    \caption{}
    \label{fig:compa_C}
\end{subfigure}
\caption{A mock spectrum (black line) is constructed with the spherical model in two versions, LR (panel a) and HR (panel b). Best fit models
using the spherical model (red line), the wind model (blue line) and the disk model (yellow line) are overplotted.  Note that far wings of the BLR emission are not considered in the fit. }
\end{figure}

Fig.~\ref{fig:compa_A} and Fig.~\ref{fig:compa_C} show the mock spectrum (black line) in LR and HR versions respectively along with the best fit for all models. It can be seen that at low resolution and low SNR, it is not possible to discriminate between the three models. However, differences appear at higher resolution and SNR.
The disk model fails to reproduce the shape of the trough. 
On the other hand, one can see that the wind model can reproduce the overall shape of the emission but fails to reproduce the far wings of the line and the two peaks of the emission differ slightly from the spherical ones. Note that the far wings are not taken into account in the fit as they may be a poor discriminant because of the simplicity of our models. On the contrary the differences seen in the shape of the peaks could be a good indicator to look at in real data providing the quality of the data is high enough.

\subsection{Wind model}

\begin{table*}
    \centering
    \begin{tabular}{c|c|c|c|c|c|c|c|c|c|c}
         \textbf{\#}&\textbf{Method} &\textbf{Quality}& \textbf{$N$(H~{\sc i})} & \textbf{Radius} & \textbf{x Coordinate}& \textbf{y Coordinate} &\textbf{Inclination}& \textbf{Strength NLR}  & \textbf{Strength BLR}& \textbf{$\chi^2$} \\
    \hline
    \hline
    
    1&Wind & N.A. & 20.3 & 0.4 & 0.16 & 0.32 & 60 & 0.6 &  1.4 & N.A\\
    \hline\hline
    2&Wind & LR & \textbf{20.3} & \textbf{0.4} & 0.08 & 0.36 & 50 & 0.80 & 1.31 & 0.86\\
    3&Wind & HR & \textbf{20.3} & 0.6 & 0.60 & 0.0 & \textbf{60} & \textbf{0.56} & \textbf{1.40} & 1.11\\
    \hline
    4&Spherical & LR & \textbf{20.3} & 0.3 & 0.30 & 0.0 & N.A. & \textbf{0.57} & 1.16 & 1.08\\
    5&Spherical & HR & \textbf{20.3} & 0.1 & 0.1 & 0.0 & N.A. & 0.37 & 1.17& 5.22\\
    \hline
    6&Disk & LR & \textbf{20.3} & 0.6 & 0.0 & -0.30 & 0 & \textbf{0.57} & 4.25 & 0.97\\
    7&Disk & HR & \textbf{20.3} & 0.7 & 0.07 & -0.56 & 0 & 0.0 & 4.28 & 3.82\\

    \end{tabular}
    \caption{Results from the fit of a mock spectrum constructed with the wind model in two versions LR and HR. The input parameters are indicated in the first row. We fit the low and high resolution spectra with the spherical, wind  and disk models to try to recover the input parameters. The parameters of the best fits for the different models are presented in rows \#2 to \#7. 
    When the fit recovers the initial parameter within 10\%, the value is printed in boldface.}
    \label{tab:from_conetore}
\end{table*}

We construct a mock spectrum using the wind model, the input parameters of which are given in the first row of Table~\ref{tab:from_conetore}.

We then fit the mock spectrum in its two versions, LR and HR, with the spherical, wind and disk models. Results of the best fits are given in rows \#2 to \#7 of Table~\ref{tab:from_conetore}.
From the $\chi^2$ values given in the table, it is clear that we cannot discriminate between the different models in LR.
But, and as for the spherical model, we can do so if high quality (HR) data are available.

An important fact is that independently of the resolution or of the model, the correct H~{\sc i} column density is recovered. However, this is not the case for the other parameters except the inclination and the BLR to NLR emission ratio for the wind model in HR.

\begin{figure}
\begin{subfigure}[b]{0.48\textwidth}
    \includegraphics[trim = 3.cm 0.cm 4.5cm 2.5cm, clip,width=\textwidth]{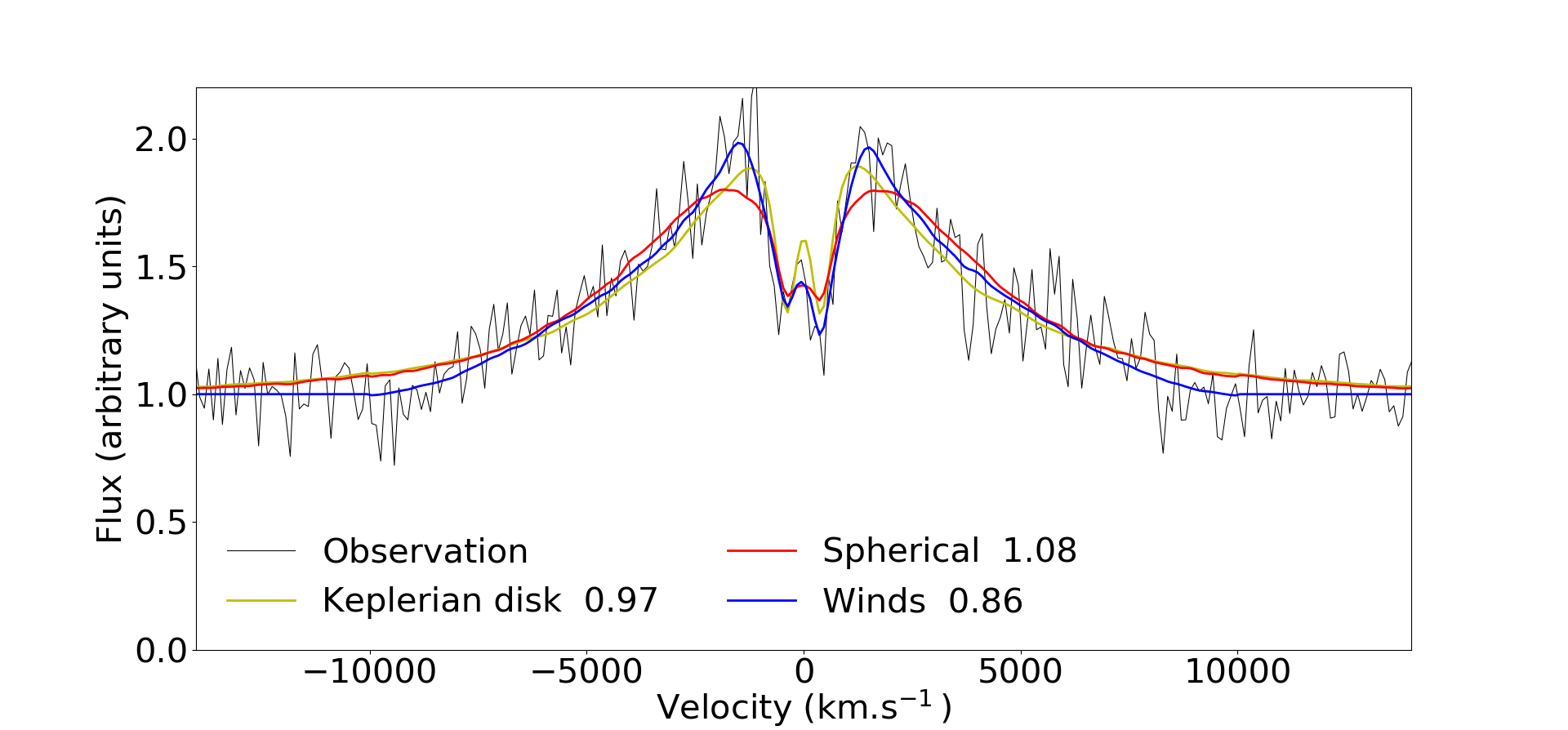}
    \caption{}
    \label{fig:compa_E}
\end{subfigure}

\begin{subfigure}[b]{0.48\textwidth}
    \includegraphics[trim = 3.cm 0.cm 4.5cm 2.5cm, clip,width=\textwidth]{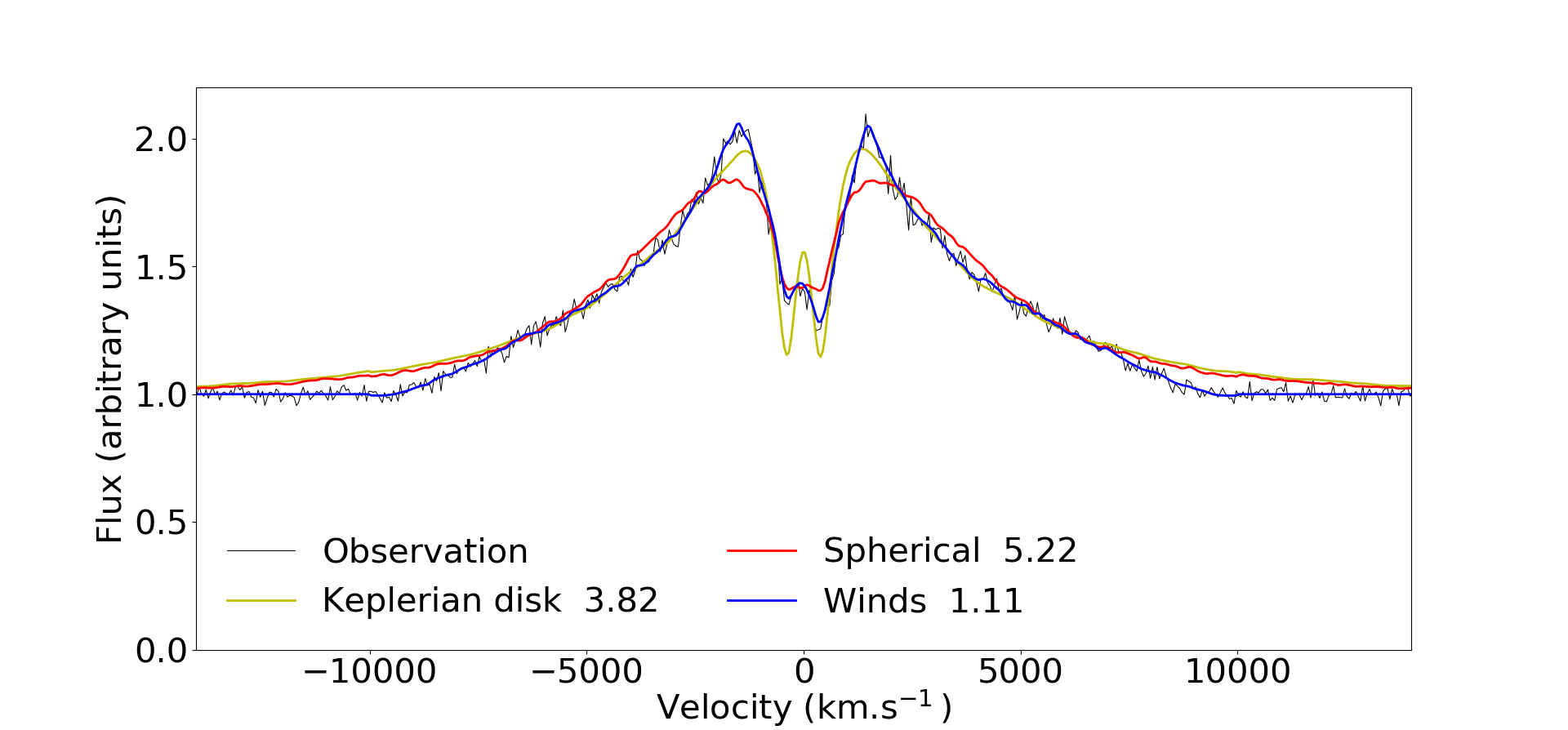}
    \caption{}
    \label{fig:compa_D}
\end{subfigure}
\caption{A mock spectrum (black line) is constructed with the wind model in two versions, LR (panel a) and HR (panel b). Best fit models
using the spherical model (red line), the wind model (blue line)  and the disk model (yellow line) are overplotted.  The far wings of the emission
are not taken into account in the fit.}
\end{figure}

Fig.~\ref{fig:compa_E} and Fig.~\ref{fig:compa_D} show the mock spectrum (black line) in LR and HR versions respectively along with the best fit for the three models. Again, at low resolution and low SNR, it is difficult to discriminate between the models even though the spherical model (red line) seems to show too flat peaks. This impression is confirmed in HR.
We observe on Fig.~\ref{fig:compa_D} that the flatness of the peaks of the spherical model does not allow this model to fit the mock spectrum well.
We can also notice that the absorption feature of the mock spectrum is asymmetric which cannot be reproduced by the spherical model. Indeed, this asymmetry is the result of the spatial structure of the BLR in the wind model. 
For the spherical model the absorption is bound to be symmetric as every cloud has the same probability to have a positive or negative velocity. In other words, negative and positive velocities are absorbed in the same way independently of the position or the size of the absorbing cloud.

The disk model spectrum shows peaks with a flatness intermediate between that of the spherical and wind models. The largest difference between the disk model and the mock spectrum, although not prominent, resides in the shape of the central part of the absorption trough which is due to the peculiar inclination of the disk.

As Table~\ref{tab:from_conetore} shows, the best inclination is 0° which means that the mean plane of the disk is perpendicular to the line of sight.

\subsection{Keplerian disk model}

\begin{table*}
    \centering
    \begin{tabular}{c|c|c|c|c|c|c|c|c|c|c}
         \textbf{\#}&\textbf{Method} &\textbf{Quality}& \textbf{$N$(H~{\sc i})} & \textbf{Radius} & \textbf{x Coordinate}& \textbf{y Coordinate} &\textbf{Inclination}& \textbf{Strength NLR}  & \textbf{Strength BLR}& \textbf{$\chi^2$} \\
    \hline
    \hline
    
    1&Disk& N.A. & 21.0 & 0.4 & 0.08 & -0.32 & 40 & 0.69 &  0.41 & N.A\\
    \hline\hline
    2&Disk & LR & \textbf{21.0} & \textbf{0.4} & 0.0 & -0.40 & \textbf{40} & 0.61 & \textbf{0.37} & 0.82\\
    3&Disk & HR & \textbf{21.0} & \textbf{0.4} & \textbf{0.08} & \textbf{-0.32} & \textbf{40} & \textbf{0.69} & \textbf{0.41} & 1.26\\
    \hline
    4&Spherical & LR & 21.2 & \textbf{0.4} & 0.12 & 0.0 & N.A. & 0.44 & 0.86 & 1.06\\
    5&Spherical & HR & \textbf{21.0} & 0.9 & 0.0 & 0.0 & N.A. & 0.83 & 0.70 & 5.67\\
    \hline
    6&Wind & LR & 21.2& 0.7 & -0.28 & 0.56 & 60 & 0.38 & 1.04 & 0.90\\
    7&Wind & HR & 21.2 & 0.9 & -0.72 & 0.09 & 70 & 0.52 & 1.40 & 2.58\\
    
    \end{tabular}
    \caption{Results from the fit of a mock spectrum constructed with the disk model in two versions LR and HR. The input parameters are indicated in the first row. We fit the low and high resolution spectra with the spherical, wind  and disk models to try to recover the input parameters. The parameters of the best fits for the different models are presented in rows \#2 to \#7. 
    When the fit recovers the initial parameter within 10\%, the value is printed in boldface.}
    \label{tab:from_KN}
\end{table*}

As we did previously, we construct a mock spectrum with the Keplerian disk model whose input parameters are displayed in the first row of Table~\ref{tab:from_KN}.
The best fit parameters for each model in LR and HR are presented in rows \#2 to \#7. 
The resulting $\chi^2$ values indicate that the fits in LR are good for all three models. These fits are significantly worse in HR for the spherical and wind models whereas the fit is good for the disk model, as expected. Once again we can recognize the model used to built the mock spectrum providing good spectral resolution and SNR are used for the observations.

We notice that the column density is pretty well recovered for all models.
Besides, in HR, the fit using the disk model retrieves all the input parameters including the BLR to NLR flux ratio, together with the size and position of the absorbing cloud which is promising.

\begin{figure}
\begin{subfigure}[b]{0.48\textwidth}
    \includegraphics[trim = 3.cm 0.cm 4.5cm 2.cm, clip,width=\textwidth]{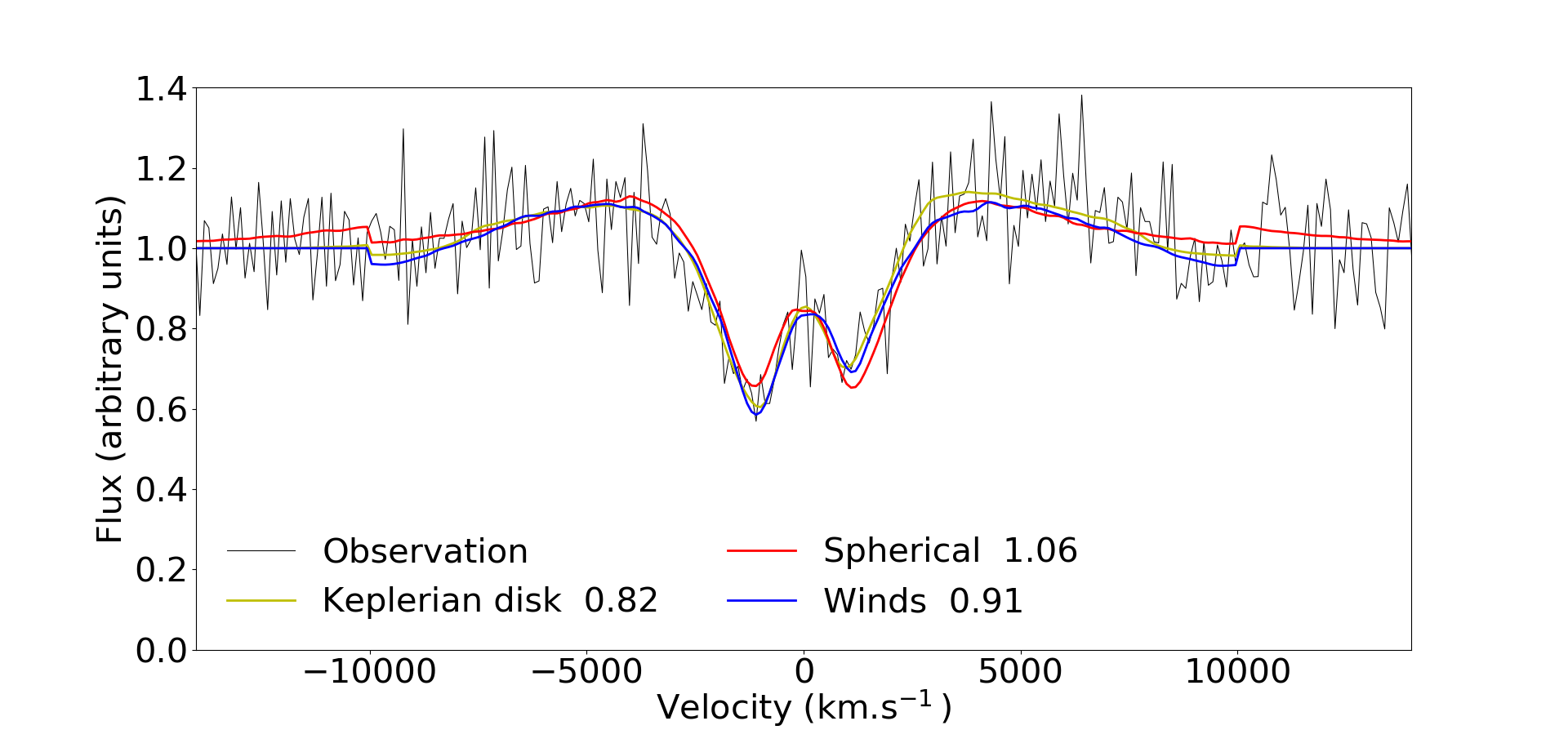}
    \caption{}
    \label{fig:compa_KN_LR}
\end{subfigure}

\begin{subfigure}[b]{0.48\textwidth}
    \includegraphics[trim = 3.cm 0.cm 4.5cm 2.cm, clip,width=\textwidth]{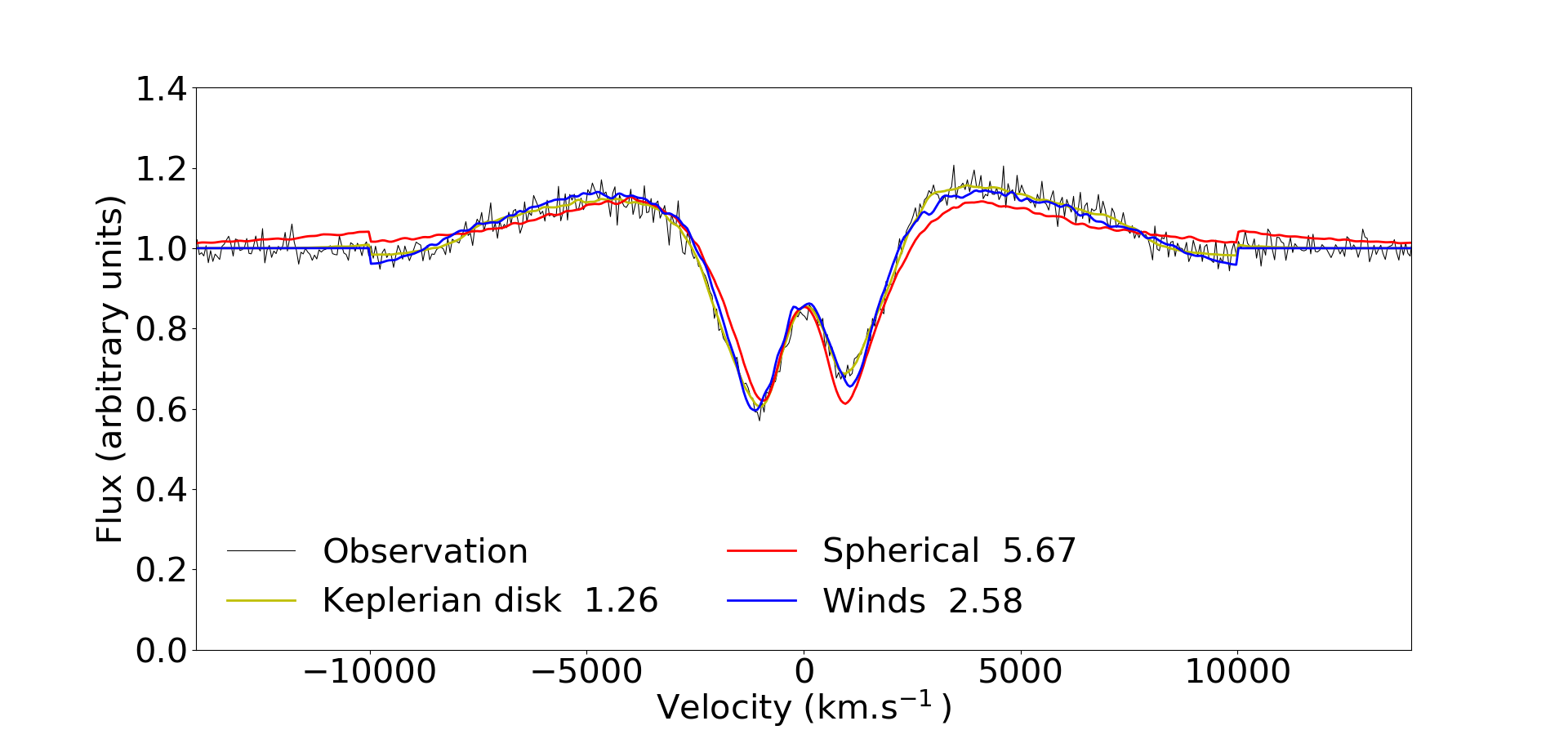}
    \caption{}
    \label{fig:compa_KN_HR}
\end{subfigure}
\caption{A mock spectrum (black line) is constructed with the Keplerian disk model in two versions, LR (panel a) and HR (panel b). Best fit models
using the spherical model (red line), the wind model (blue line) and the disk model (yellow line) are overplotted. Note that the far wings of the emission are not taken into account in the fit.
}
\end{figure}

On Fig.~\ref{fig:compa_KN_LR} and Fig.~\ref{fig:compa_KN_HR}, we display the mock spectrum together with the best fits using the three models. We can see that even in LR, the spherical model struggles to fit the mock spectrum properly as it fails to reproduce the asymmetry of the trough. This is even more apparent in HR. On the other hand, the wind model succeeds to reproduce the overall shape of the line in LR and even in HR except for some minor features.

\subsection{Summary}

From this study, we find that the model used to produce the mock spectrum can be recovered for the three models. However this is valid only if the spectral resolution and the SNR are sufficient, meaning that with the quality of SDSS data, the distinction will be difficult.
With higher quality data, some specific features can be used to discriminate between models such as the asymmetry of the trough rejecting automatically the spherical model. The flatness of the spherical model peaks is also a good indicator. In general, the wind and disk models are more versatile and are more difficult to disentangle.

An important result is that even with LR data, we can derive a good estimate of the H~{\sc i} column density. This will be investigated in more details in the next section.

\section{Investigations of the parameters}

We have shown in the previous section that we can reproduce ghostly-DLAs well and that there is a promising difference between models. In this section we will try to understand if, {\sl once a model is preferred}, we can derive quantitative constraints on parameters such as the inclination of the BLR in the case of the wind and disk models and the size, position and column density of the absorbing cloud.

To estimate these constraints, we first fix a reduced $\chi^2$ threshold value to define an acceptable fit. Even though a $\chi^2$ closer to 1 is considered better, a rule of thumb states that a value below $1.5$ indicates an acceptable fit. 
It does not mean that a fit above this limit is bad. It only gives us a way to compare $\chi^2$ values between the different fits.

We use the same mock spectra built from the three models with parameters as given in the first row of Tables~\ref{tab:from_spherical}, \ref{tab:from_conetore} and \ref{tab:from_KN} and we fit the mock spectra with the best model as derived from the previous section.

In the following, we chose one input parameter, fix its value and vary all other parameters deriving the minimum $\chi^2$. We finally vary the value of the chosen parameter and study the evolution of this minimum $\chi^2$.

\subsection{Column density of the absorbing cloud}

As said before, when the redshift of the absorber is high enough, the DLA column density can be inferred from the absorption lines of the Lyman series. However, in most cases, only the Ly$\alpha$ line wavelength range is available.

\begin{figure}
    \centering
    \includegraphics[trim = 3.5cm 0.cm 3.8cm 2cm clip,width=0.48\textwidth]{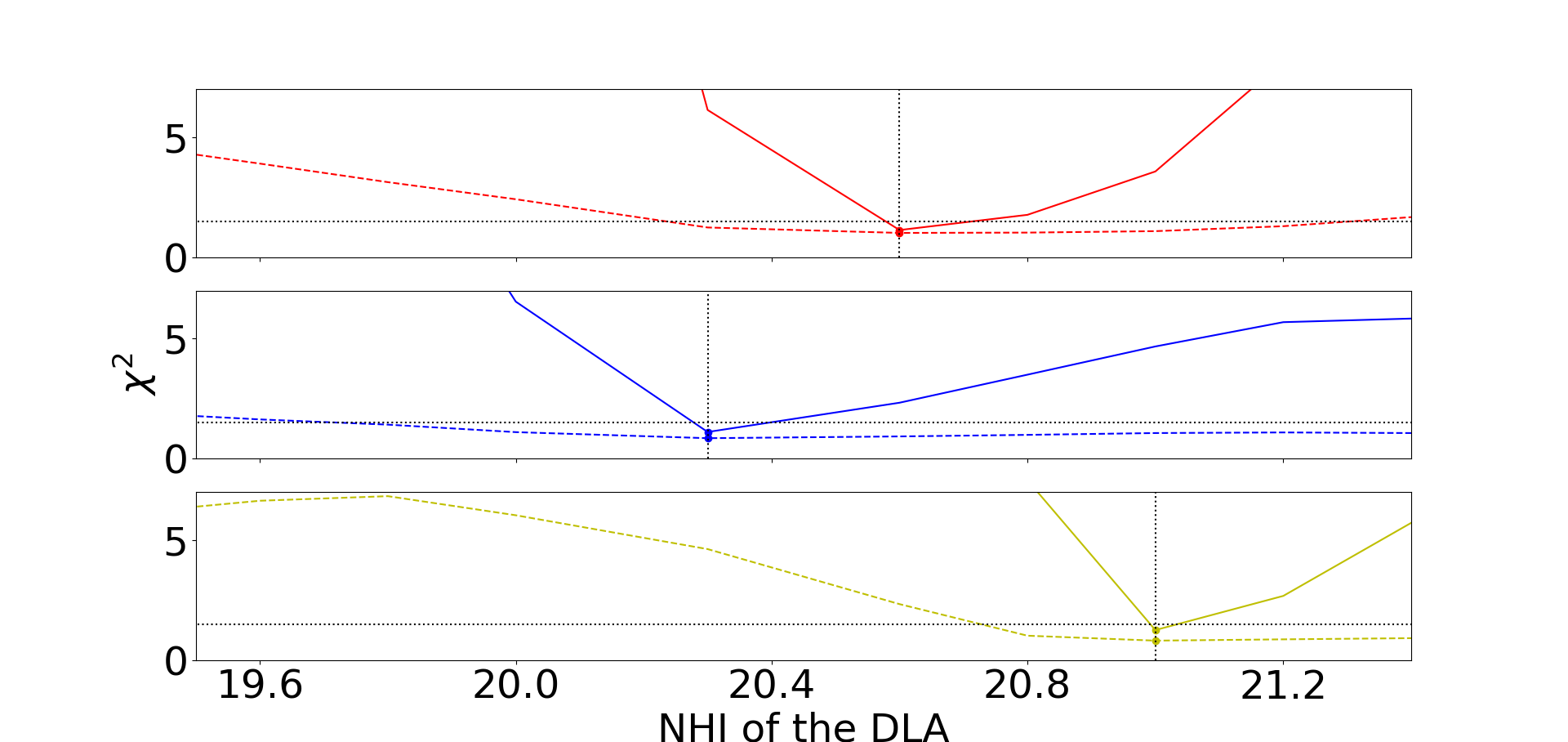}
    \caption{Minimum $\chi^2$ as a function of the DLA column density for the different mock spectra obtained, from top to bottom panel, using the spherical, the wind and the Keplerian disk model respectively. In each panel, the minimum $\chi^2$ evolution is plotted as a dashed line (resp. solid line) for spectra in LR (resp. HR).
    The black vertical dotted lines correspond to the input column densities. 
    The thin black horizontal dotted lines correspond to the $\chi^2=1.5 $ threshold.
    }
    \label{fig:density}
\end{figure}

Fig.~\ref{fig:density}, shows the minimum $\chi^2$ as a function of the neutral column density for the three models in LR and HR. 
The minimum of each curve is indicated by a colored dot. 

In the three cases, the correct $N$(H~{\sc i}) value indicated by a vertical line is retrieved by the model which the mock spectrum is based on. The determination is more precise with high quality data (HR).

\subsection{Inclination of the BLR}

\begin{figure}
    \centering
    \includegraphics[trim = 3.5cm 0.cm 3.8cm 2cm clip,width=0.49\textwidth]{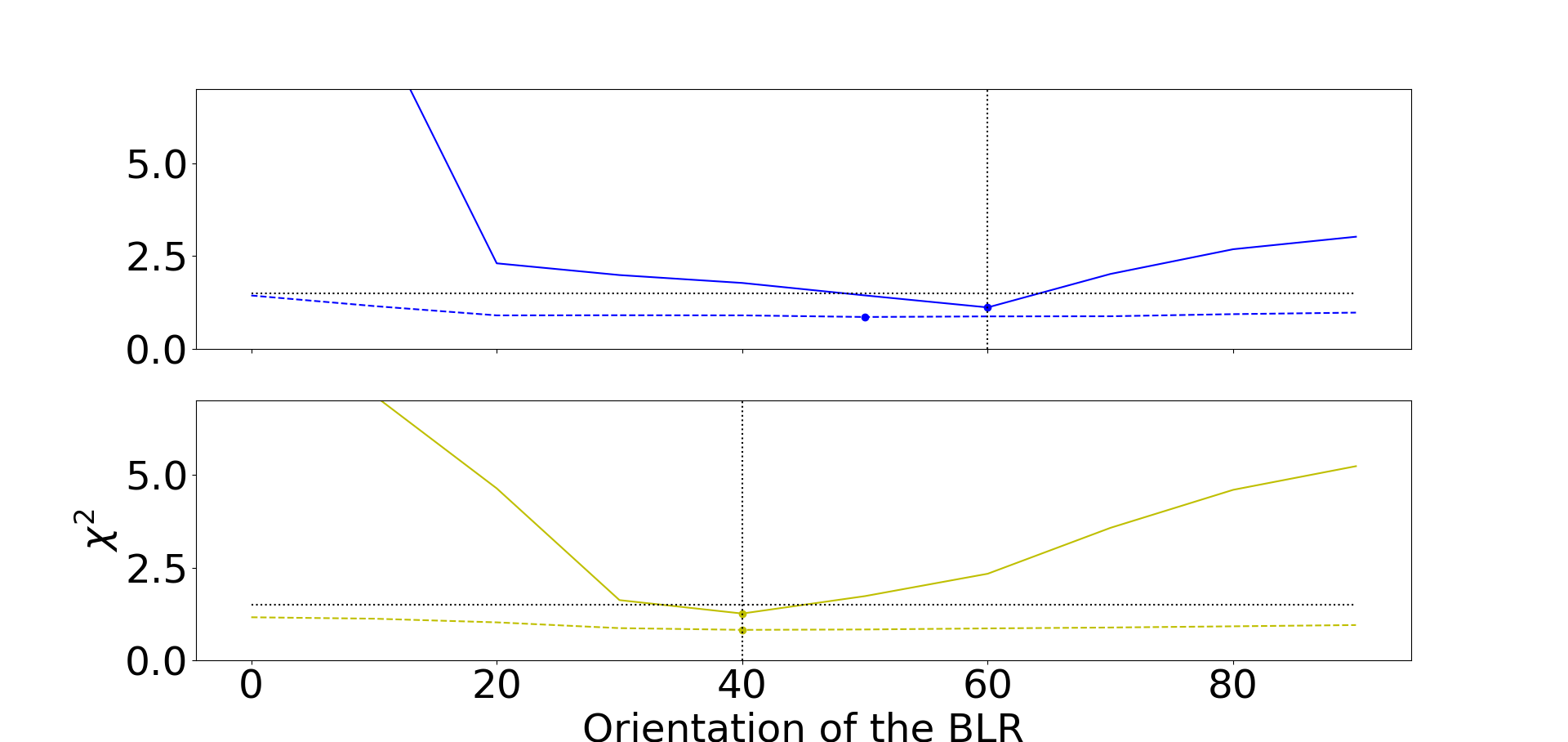}
    \caption{Minimum $\chi^2$ as a function of the inclination of the BLR with respect to the observer for the wind model (\textit{top panel}) and the disk model (\textit{lower panel}) mock spectra. On each panel, the minimum $\chi^2$ evolution is plotted as a dashed line (resp. solid line) for spectra in LR (resp. HR). 
    The black vertical dotted lines correspond to the input inclination.
    The thin black horizontal dotted lines correspond to the $\chi^2=1.5 $ threshold.
    }
    \label{fig:angle}
\end{figure}

Fig.~\ref{fig:angle} represents the minimum $\chi^2$ as a function of the inclination angle with respect to the observer for the wind and the disk mock spectra. We do not use the spherical model as the later is symmetric and has no inclination parameter.

The minimum value of each curve is indicated by a colored dot. We observe that the two curves have a minimum at the correct inclination of the BLR of their respective mock spectrum. However, the constraints are weak.
Using the $\chi^2$ threshold given before, we cannot constrain the inclination when fitting LR spectra.With HR spectra, the inclination is better constrained.

From this comparison, it is again clear that high quality data are needed 
to constrain this parameter within a decent error box.

\subsection{Size and position of the absorbing cloud}

It is easy to foresee that the position and size of the absorbing cloud are degenerated parameters. 
The reason for this is that the emitting cloud density in the BLR is decreasing outwards. Two absorbing clouds with different radius can yield a similar spectrum provided the largest one is located further away from the center because it will cover a larger but less dense region.

This is why for the three mock spectra with both resolutions, the minimum $\chi^2$ as a function of the radius is almost constant and no clear minimum is seen.

\begin{figure}
\includegraphics[trim = 3.cm 0.cm 4.5cm 2.5cm, clip,width=0.5\textwidth]{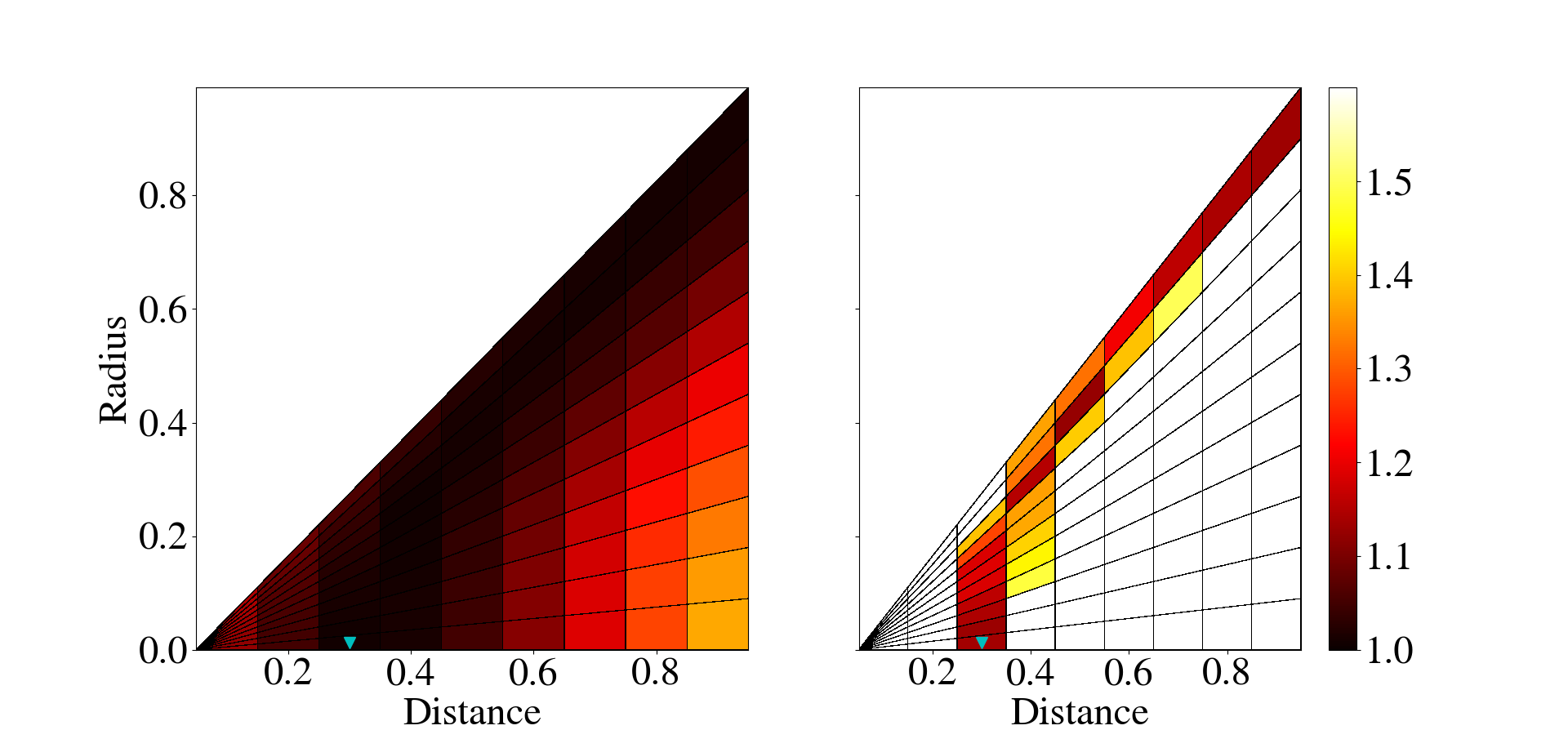}
\caption{Maps representing the $\chi^2$ as a function of the radius of the cloud and its distance to the centre for the  mock spectra obtained using the spherical model in LR (\textit{left panel}) and HR (\textit{right panel}). The blue triangle indicates the radius and distance used to obtain the mock spectrum. The $\chi^2$ color scale is shown on the right hand side of the figure. }
\label{fig:map_s}
\end{figure}

Here again, we will fit the mock spectra with the model that has been used to construct it.
The correlation between the cloud radius and its distance to the centre is illustrated on Fig.~\ref{fig:map_s}, showing the $\chi^2$ as a function of those two parameters  when using the spherical model. 
It is interesting to note that with high quality data it is possible to derive a lower limit of the radius, because the cloud must in any case cover the central region where the quasar continuum is emitted. In addition, the correlation between the distance to the centre and the radius of the cloud is tight which means that the cloud cannot be much larger than its distance to the centre. This is a very interesting constraint as, one could estimate independently the radius of the cloud by deriving the particle density in case C~{\sc i} absorption lines are detected \citep{Fathivavsari_2017}.

\begin{figure}
    \centering
    \includegraphics[trim = 11cm 2cm 7cm 2.3cm, clip,width=0.5\textwidth]{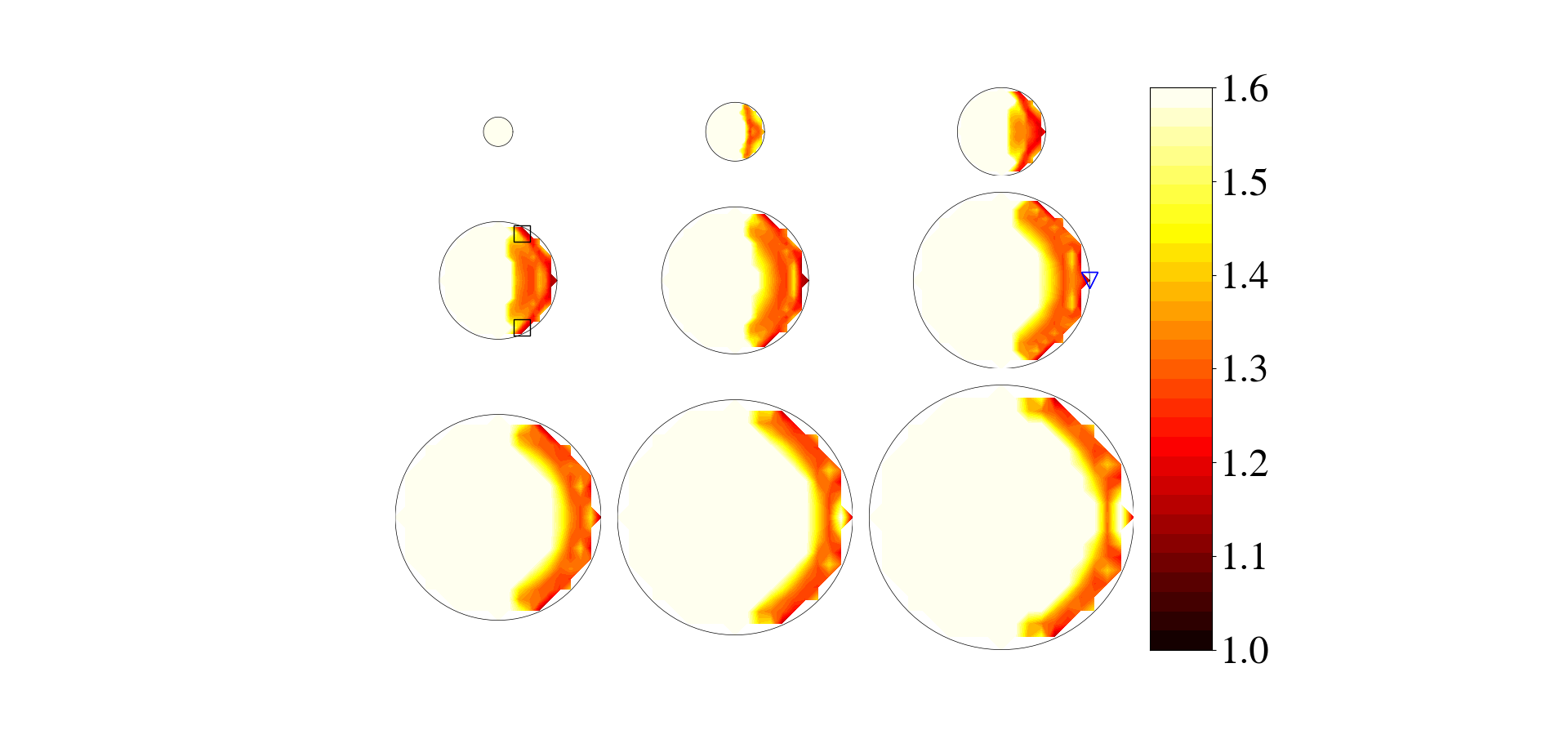}
    \caption{$\chi^2$ as a function of the position and radius of the cloud in HR for the wind model. The best fit position is indicated by a blue triangle whereas the two possible input positions are indicated by two black boxes. There are two possible input positions due to the  symmetry relative to the x-axis.}
    \label{fig:map_c}
\end{figure}

For the wind and disk models, the $\chi^2$ does not depend only on the radius and distance to the centre but also on the exact position of the cloud therefore its x and y coordinates.
Fig.~\ref{fig:map_c} shows the $\chi^2$ at each position of the cloud for a radius varying from 0.1 on the left-top corner to 0.9 on the right-bottom corner with an increment of 0.1 when fitting the HR wind mock spectrum. 
In LR, almost all positions of the absorbing cloud give a good fit and it is possible to constrain neither the size nor the position. That is why only the HR version is discussed here.
It is apparent that the radius and the distance to the centre are degenerated because tightly correlated for this model as well. For each cloud radius, the best fit is 
obtained with a cloud  at a distance corresponding roughly to the radius. 
One can also notice that the right side of the BLR is clearly favored in the fit. This is due to the asymmetry of the absorption which favors one side of the model. 
Note that the model being symmetric relative to the x-axis, there are two possible input positions for the same mock spectrum.

\begin{figure}
    \centering
    \includegraphics[trim = 11cm 2cm 7cm 2.3cm, clip,width=0.5\textwidth]{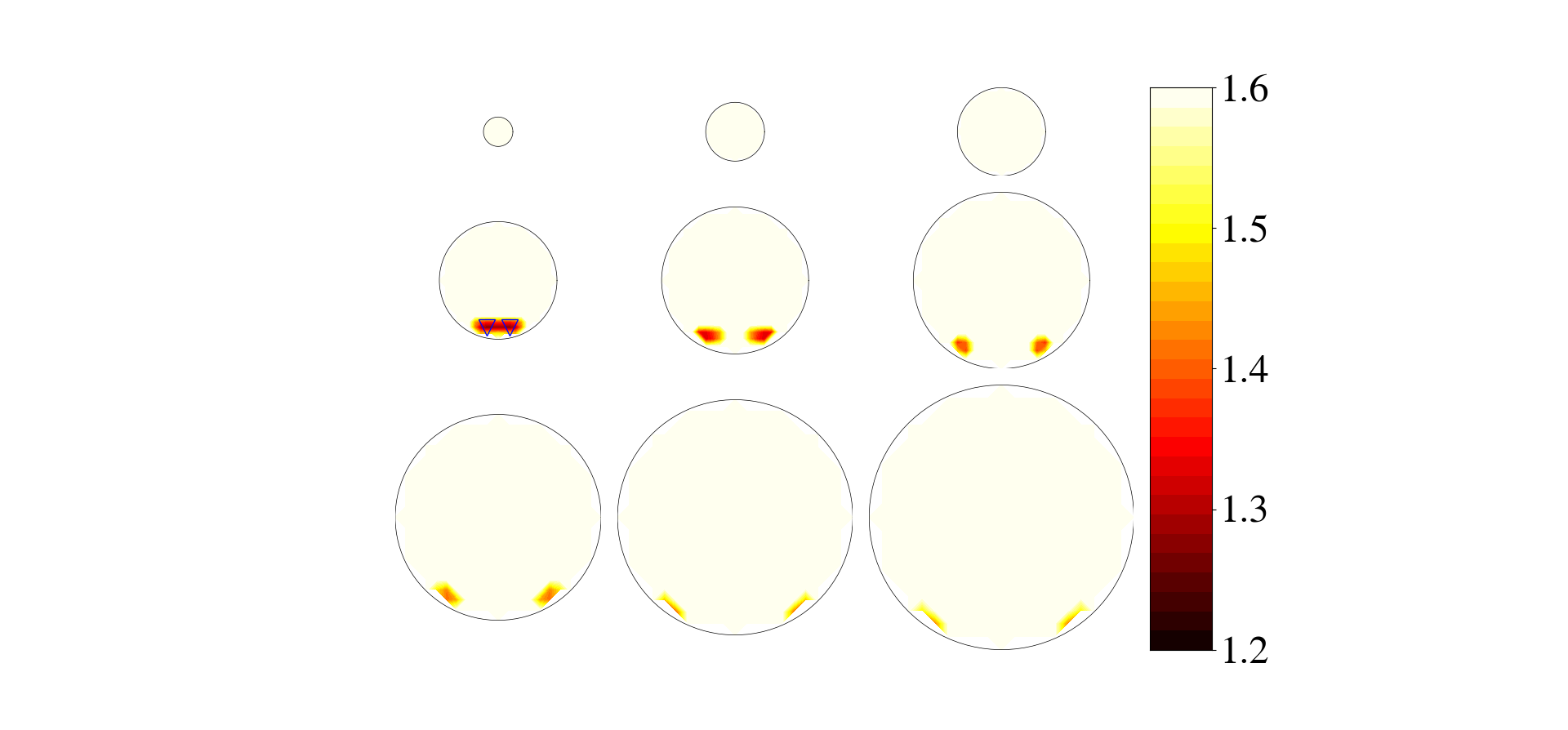}
    \caption{$\chi^2$ as a function of the position and radius of the cloud in HR for the disk model. The best fit positions, which are also the input positions, are indicated by two blue triangles. There are two possible input positions due to the vertical symmetry.}
    \label{fig:map_KN}
\end{figure}

\begin{figure}
    \centering
    \includegraphics[trim = 5.5cm 1cm 8cm 1.cm, clip,width=0.5\textwidth]{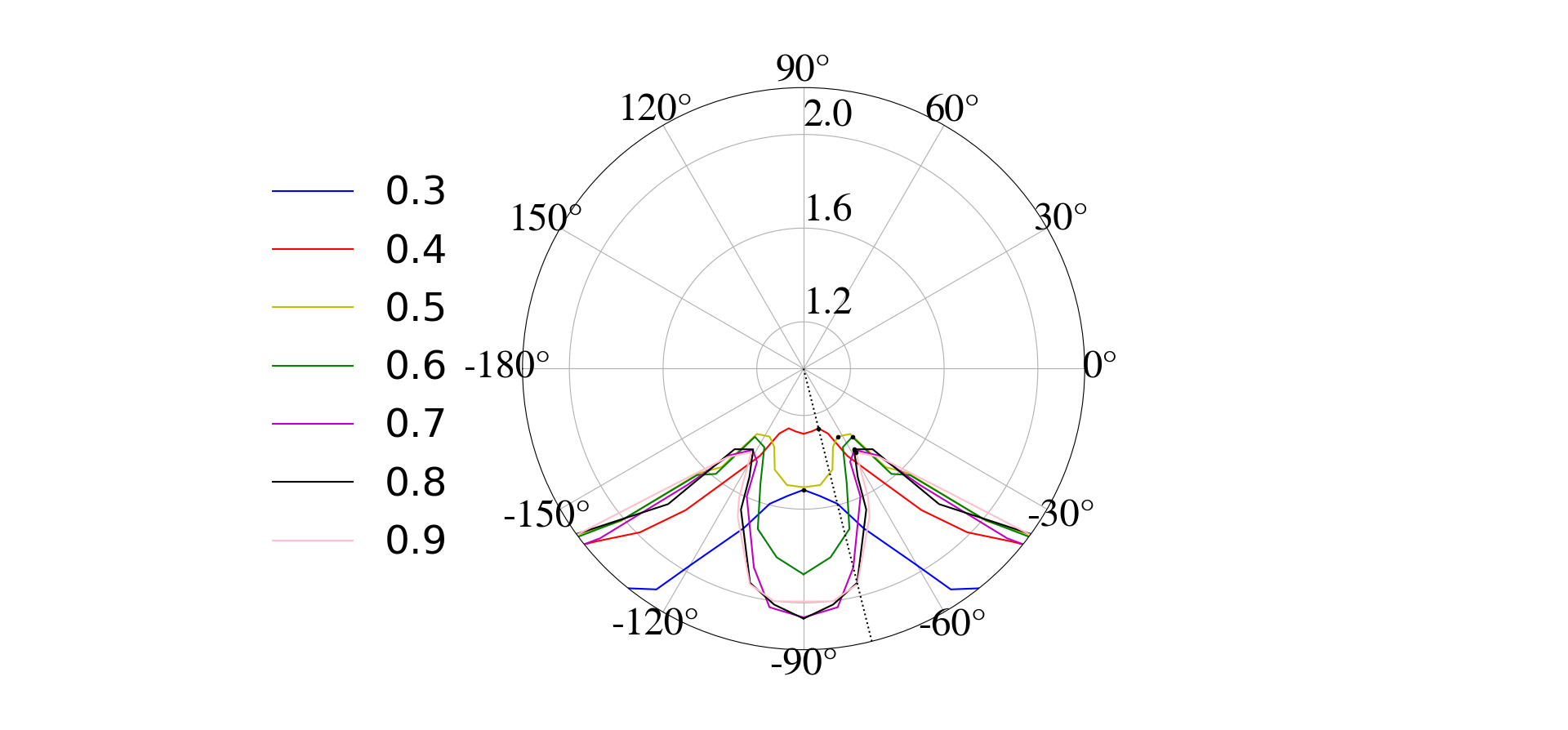}
    \caption{Curves ($\chi^2$, $\phi$) showing, for the wind model and each absorbing cloud radius, the minimum $\chi^2$ found along the radial direction defined as the angle, $\phi$, of the position of the cloud centre relative to the y-axis.
    The color scale of the cloud radius is given on the left hand side of the figure.}
    \label{fig:direction}
\end{figure}

For the disk model, we observe on Fig.~\ref{fig:map_KN} a similar effect but even more apparent. Indeed, the locus of parameters yielding good fits is smaller and the radius can be constrained between 0.4 and 0.7 when the correct value is 0.4. 
We can also notice that the direction is pretty well constrained.
This is investigated further on Fig.~\ref{fig:direction}. This plot represents the minimum $\chi^2$ as a function of the direction $\phi$ of the absorbing cloud with respect to the y-axis for different values of the radius.
The minimum is reached with a direction of approximately -76° which is the correct value represented by a black dotted line. At the other radius, the figure tends to show a preferred direction of -60°. 
This clearly shows that a preferred direction can be derived especially if the radius can be constrained independently from estimating the density using C~{\sc i} lines.

\subsection{Summary}

To summarize the results of the above exercise, we can conclude the following within the framework of the three models described above.
(i) Not unexpectedly, high quality data are preferred in all cases.
(ii) The neutral hydrogen column density of  the absorbing cloud can be estimated independently of the model used with reasonable precision even with LR data.
(iii) The radius and position of the absorbing cloud are degenerated. However, if the radius of the cloud can be estimated by an independent method, then the position of the cloud can be constrained. 
(iv) The radius of the cloud has to be larger but not
much larger than the distance to the centre. 
For all other parameters, constraints are not strong and may be possible only if the best of the three models can be determined unambiguously.

\section{Fit of SDSS spectra}

In this section, we compare our models with observational data from the Sloan Digital Sky Survey (SDSS) data release 12. 
We use the sample of \textit{ghostly}-DLAs listed by \citet{Fathivavsari_2020}.
By definition a \textit{ghostly}-DLA is characterized by the presence of strong metal lines whereas the expected corresponding strong H~{\sc i} Lyman-$\alpha$ trough is not seen in the quasar spectrum. 
In some cases, no trace of the H~{\sc i} absorption can be seen.  In that case, it is not possible, without any additional information to constrain our models.
In other cases however, some residual of the H~{\sc i} absorption is left in the spectrum providing a direct access to the H~{\sc i} column density.  
Among the 23 \textit{ghostly}-DLAs in the sample, 7 show some absorption residual. 
Among the 7 spectra only three have a high enough signal-to-noise ratio (SNR~$>$~10) to perform a realistic fit with our models, QSO J000958.66+015755.18 having the highest SNR (SNR$>$20).

\subsection{Fit of QSO J000958.66+015755.18}

The spectrum of QSO J000958.66+015755.18 has the highest SNR in the sample of \textit{ghostly}-DLAs.
In addition, the redshift of the quasar is $z_{\rm QSO}=2.973$ which means that the Lyman-$\beta$ line is seen in the spectrum.
The \textit{ghostly}-DLA is at $z_{\rm DLA}=2.97635$, derived from the numerous strong metal absorption lines.

Before comparing the observations with our different models, we have to remove the N~{\sc v} contribution from the quasar emission. We fit a Gaussian emission located at $\lambda_{\rm rest}$~=~1240.1\AA~ and remove it from the spectrum.
During the fit we have excluded the pixels affected by strong absorptions unrelated to our system and located around $-4000$, $-2000$ and $6000$~km~s$^{-1}$.

\begin{figure}
    \centering
    \includegraphics[trim = 3.5cm 0cm 4.5cm 1.8cm, clip,width=0.45\textwidth]{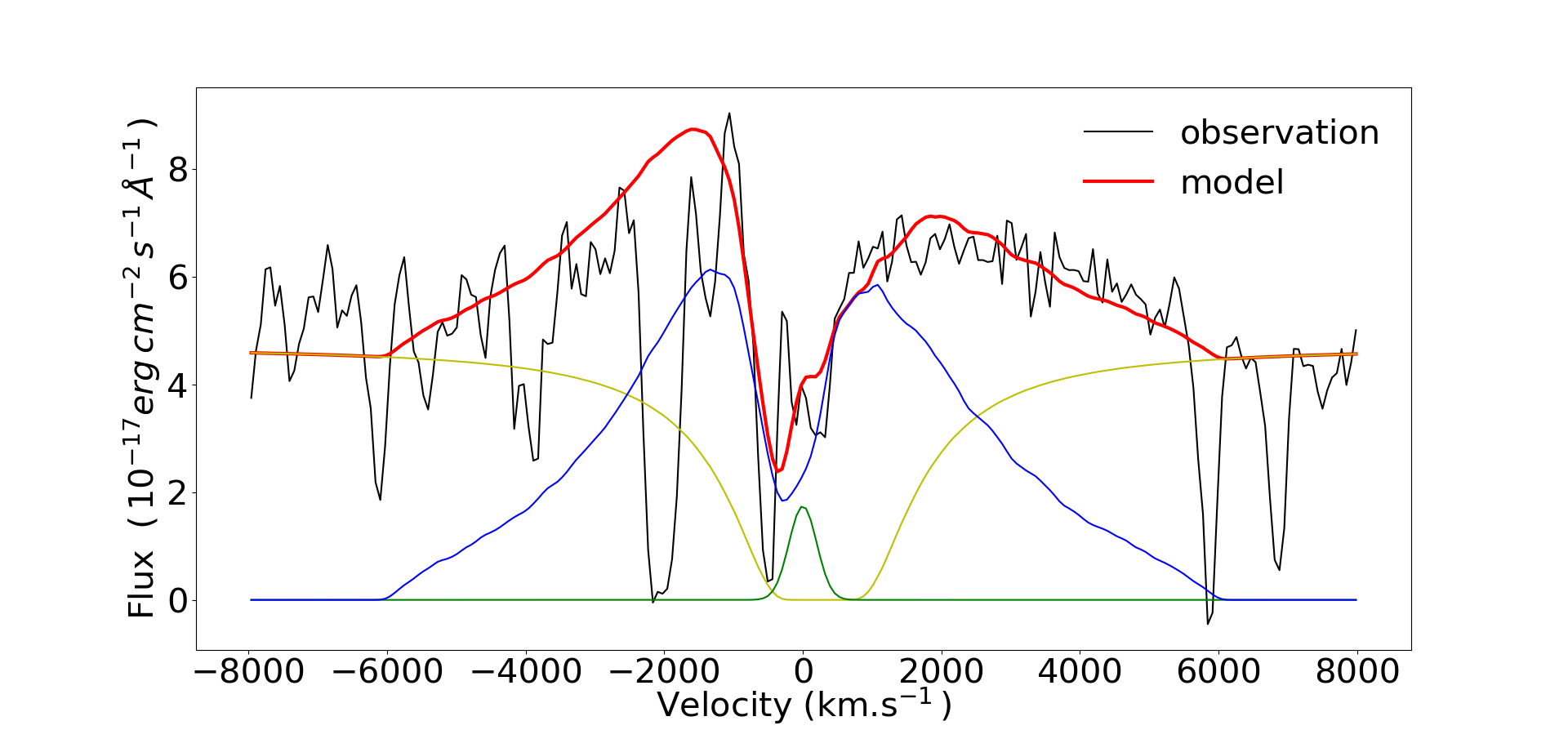}
    \caption{Best fit (red curve) of the Lyman-$\alpha$ emission of quasar J000958.66+015755.18 (black curve) with a wind model. The continuum, the BLR and NLR emissions are respectively the yellow, blue and green curves.
    }
    \label{fig:fit_obs_7}
\end{figure}
\begin{figure}
    \centering
    \includegraphics[trim = 0.4cm 1.5cm 1.2cm 1.8cm, clip,width=0.3\textwidth]{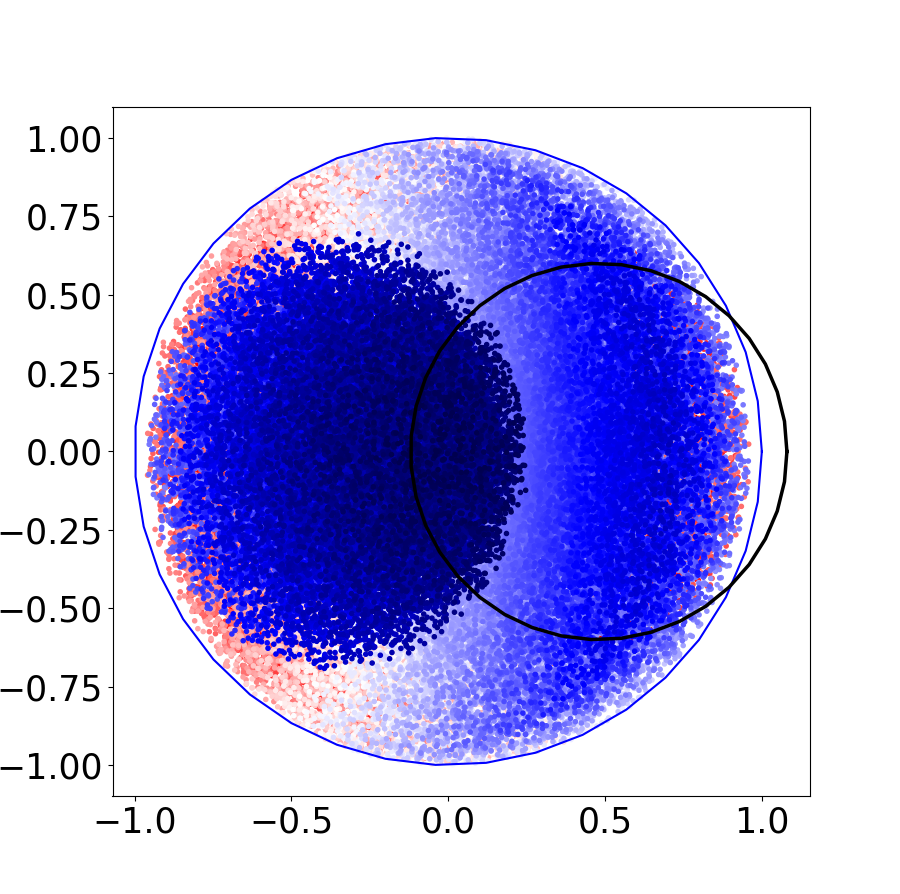}
    \caption{Location of the BLR emitting clouds colored as a function of their relative velocity for the best model of
    J000958.66+015755.18. The model is a wind model with inclination $i=30$°.
    The absorbing cloud with radius $r=0.6$
    at location (0.4,0) is indicated by the black circle.}
    \label{fig:fit_obs_7_dispo}
\end{figure}

The best fit of the QSO Lyman-$\alpha$ emission line is displayed on Fig.~\ref{fig:fit_obs_7}. The fit is a wind model with an inclination relative to the line of sight of 30°.
The absorption cloud has a column density of log $N$(H~{\sc i})~=~20.8, a radius $r=0.6$ and is located at $(0.48,\, 0.0)$ (see Fig.~\ref{fig:fit_obs_7_dispo}). 
The redshift of the quasar studied here is high enough so that the Lyman-$\beta$ absorption from the DLA is redshifted in the observed window. We therefore can use this line to confirm  some of our findings.
We use the best fit of the Lyman-$\alpha$ line and translate the model to Lyman-$\beta$.

For this, we adjust the continuum and add the Lyman-$\beta$ and O~{\sc vi}$\lambda\lambda$1031.9,1037.6 doublet emissions.
To model the Lyman-$\beta$ emission, we consider the exact same emission clouds in the BLR as for Lyman-$\alpha$. In addition, we assume that they also emit the O~{\sc vi} emission, i.e., the BLR has the same configuration for both species.
\citet{2002ApJ...565..773T} indicates that the Lyman-$\beta$ and O~{\sc vi} blend has a flux of approximately 0.2 that of the Lyman-$\alpha$ one. In \citet{2020ApJ...890L..28S}, the authors present  a quasar spectrum where the Lyman-$\beta$ and O~{\sc vi} doublet emissions are not blended and have a flux equal to $2.5\%$ and twice $7.5\%$ of the Lyman-$\alpha$ emission, respectively. We use the latter numbers.

The only parameter which remains unknown, is the Lyman-$\beta$/Lyman-$\alpha$ flux ratio for the NLR.
This ratio can vary between $1/3$ and $1/30$. To have an upper limit on the Lyman-$\beta$ emission, we use a ratio of $1/3$.

\begin{figure}
    \centering
    \includegraphics[trim = 3.7cm 0cm 4.5cm 2.7cm, clip,width=0.45\textwidth]{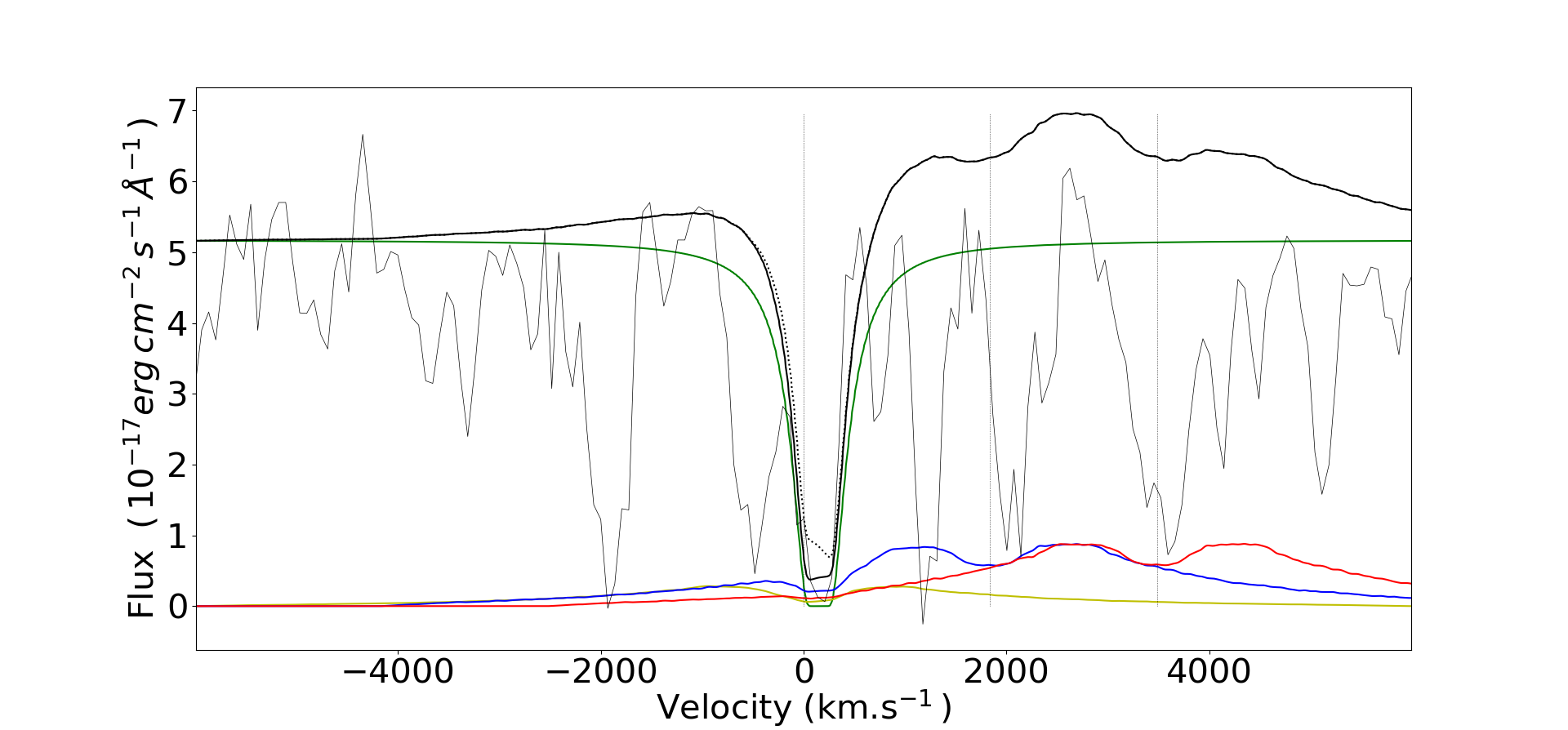}
    \caption{ We extrapolate our model to Lyman-$\beta$ with parameters constrained by the fit of the Lyman-$\alpha$ emission in quasar J000958.66+015755.18.
    The contributions of  Lyman-$\beta$, 
    and O~{\sc vi} $\lambda\lambda$1031.9,1037.6 are respectively the yellow, blue and red solid lines 
    The continuum is the green curve and the black solid line is the sum of all emission components. 
    In gray, the spectrum of the observed QSO J000958.66+015755.18. The three vertical lines indicate the positions of the Lyman-$\beta$ and O~{\sc vi} emission lines.}
    \label{fig:fit_obs_7_LyB}
\end{figure}

The result is presented on Fig.~\ref{fig:fit_obs_7_LyB}. 
The weakness of the line emission compared to the continuum explains easily why the absorption due to the DLA is detected in Lyman-$\beta$ whereas it is not detected in Lyman-$\alpha$. 
Note that the fit is good enough to confirm the H~{\sc i} column density derived from the fit of Lyman-$\alpha$ only.
It can be seen that there is some flux residual at the bottom of the Lyman-$\beta$ trough. With the quality of the SDSS data it is not possible to derive anything from it. However, using much better quality data (e.g. from XSHOOTER on the VLT), it would be possible to constrain better our model and especially the Lyman-$\alpha$/Lyman$\beta$ emission ratios.

It is possible that the absorbing cloud bears some O~{\sc vi} that could absorb the BLR O~{\sc vi} emission and the quasar continuum.
Due to the fact that Lyman-$\beta$ is located in the Lyman forest, it is difficult to test this possibility but again better quality data at higher spectral resolution could probably probe the presence of O~{\sc vi} in the cloud.

\subsection{$N$(H~{\sc i}) column densities}

We selected 2 additional quasars the spectrum of which is good enough to try to fit the Lyman-$\alpha$ emission in order to derive a neutral hydrogen column density in the cloud. Here, we briefly present the fit of these quasar spectra.

\begin{figure}
    \centering
    \includegraphics[trim = 3.8cm 0.cm 4.5cm 2.5cm, clip,width=0.45\textwidth]{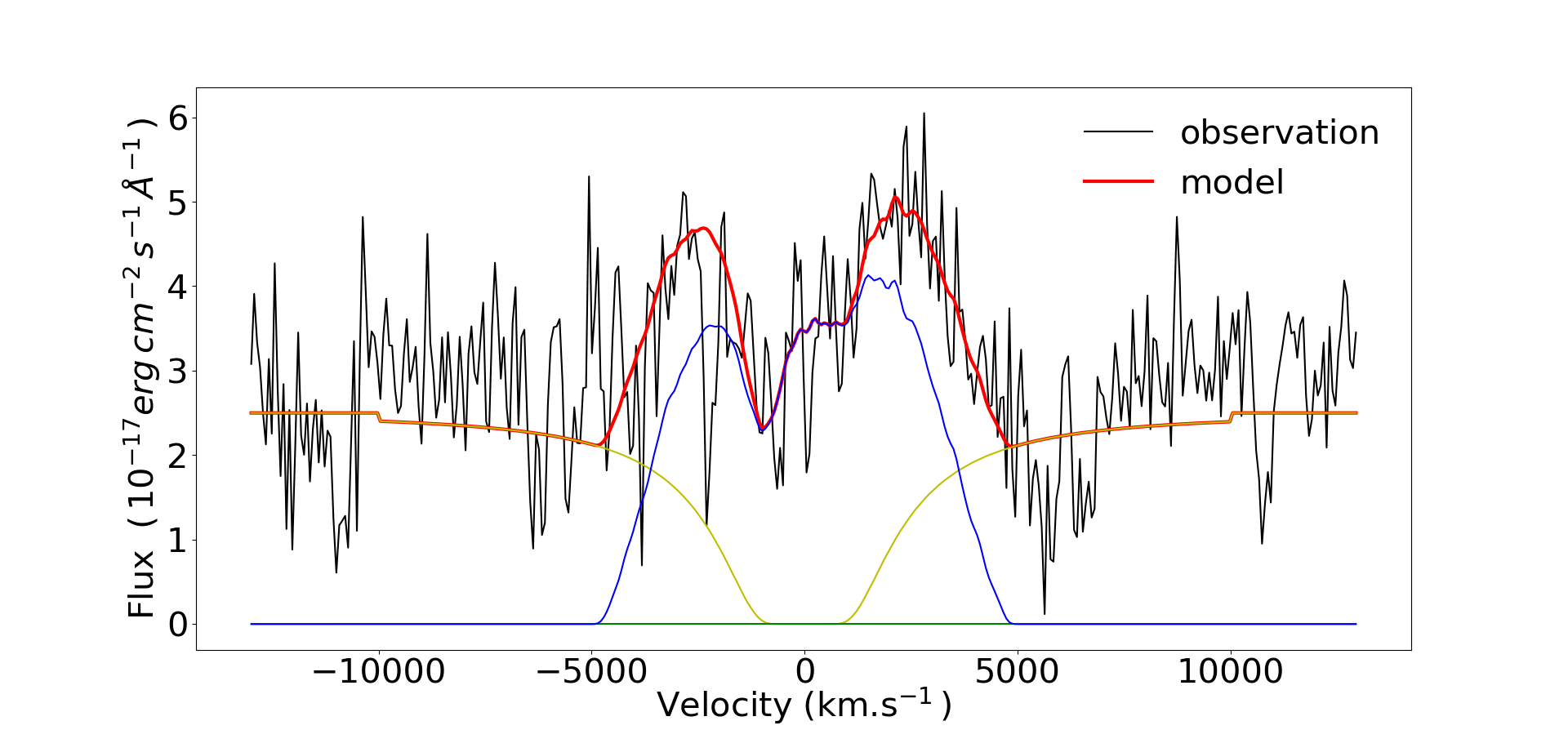}
    \caption{
    Best fit of the Lyman-$\alpha$ emission in the spectrum of QSO J124202.03-002209.00 (red curve) overplotted onto the data (black curve). 
    The continuum, BLR and NLR emissions are represented by, respectively, the yellow, blue and green curves.
    }
    \label{fig:fit_obs_2}
\end{figure}

QSO J124202.03-002209.00 has $z_{\rm QSO}=2.37925$ and $z_{\rm DLA}=2.3792$. The fit displayed on Fig.~\ref{fig:fit_obs_2} shows that no narrow component is needed to reproduce the spectrum. 
The fit is a wind model with a 60° inclination and log~$N$(H~{\sc i})~=~$21.2$.

\begin{figure}
    \centering
    \includegraphics[trim = 3.8cm 0.cm 4.5cm 2.5cm, clip,width=0.45\textwidth]{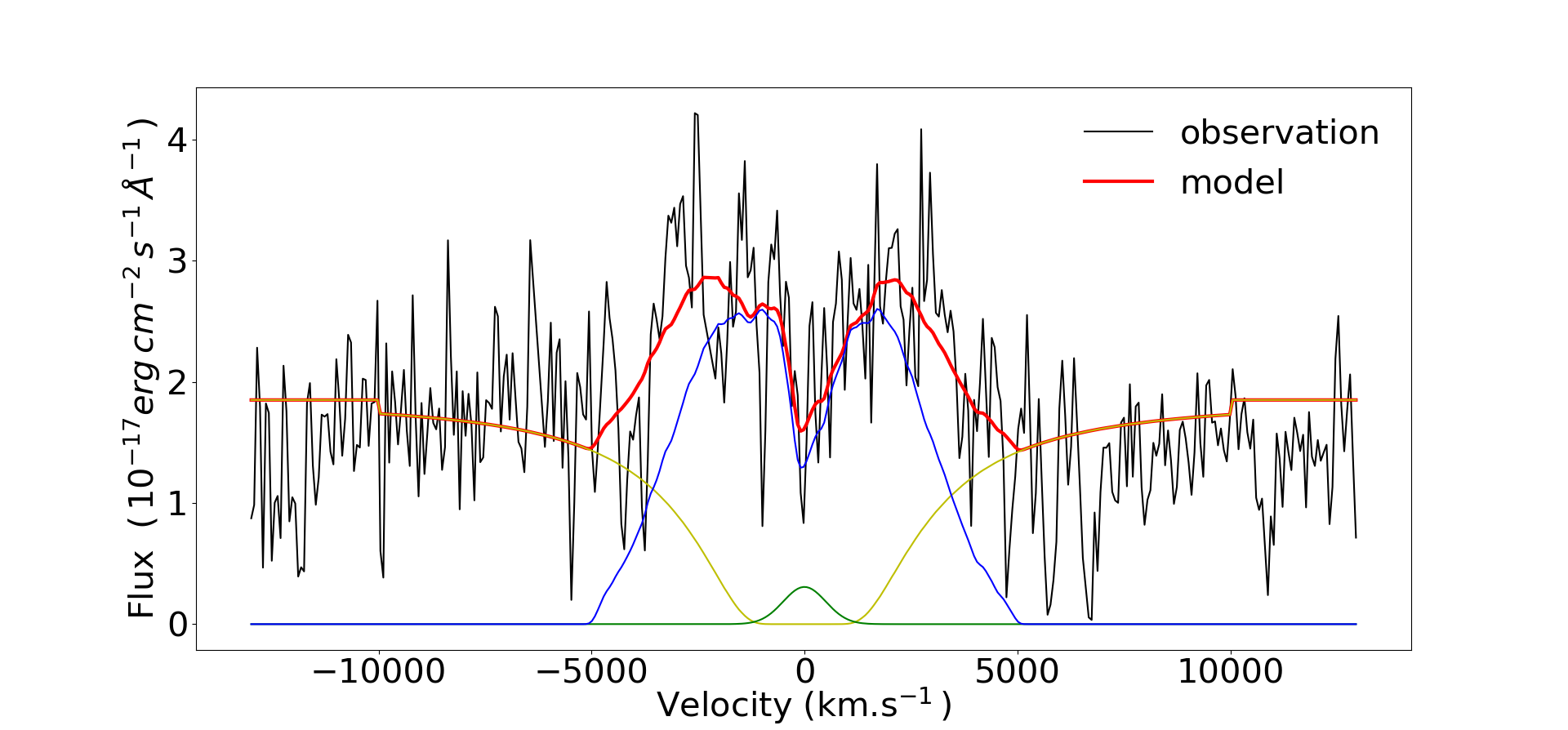}
    \caption{
    Best fit of the Lyman-$\alpha$ emission in the spectrum of QSO J125437.96+315530.84 (red curve) overplotted onto the data (black curve). 
    The continuum, BLR and NLR emissions are represented by, respectively, the yellow, blue and green curves.
    }
    \label{fig:fit_obs_6}
\end{figure}

QSO J125437.96+315530.84 has $z_{\rm QSO}=2.299$ and $z_{\rm DLA}=2.301$. The fit is presented on Fig.~\ref{fig:fit_obs_6} and one can see that a narrow component is needed but its contribution is weak. The fit is a wind model with an inclination of 40° and log~$N$(H~{\sc i})~=~$21.4$. 

For the three  \textit{ghostly}-DLAs with good enough data in SDSS, we derive
H~{\sc i} column densities of 20.8, 21.2 and 21.4. Although the number is small this shows that indeed,  \textit{ghostly}-DLAs are \textit{bona fide} damped Lyman-$\alpha$ systems with log~$N$(H~{\sc i})$>$20.3.
We note also that the three quasars are best fitted with the wind model.

\section{Conclusion}

We have constructed three geometrical models for the quasar BLR spatial and kinematical structures. The three models can reproduce the typical shape of the quasar Lyman-$\alpha$ emission.
Adding an absorption cloud in front of the BLR, we have used these models to obtain mock spectra of so-called \textit{ghostly}-DLAs. 
These absorbers are characterized by the presence of strong metal lines but no Lyman-$\alpha$ trough is seen in the quasar spectrum indicating that although the region emitting the continuum is covered by the absorbing cloud, the BLR is only partially covered.
We generate mock spectra with similar characteristics as good SDSS data ($SNR$~=~10 and spectral resolution $R$~=~2,500) but also with higher SNR and spectral resolution, $SNR$~=~50 and $R$~=~5,000.

We then try to recover the initial parameters by fitting the mock data.
We show that the H~{\sc i} column density can be recovered precisely even in SDSS data. The size of the absorbing cloud and the distance to the centre are  correlated and thus impossible to disentangle without any additional information. Only a minimal radius can be determined.

By comparing our models to SDSS data of observed \textit{ghostly}-DLAs, we show that the H~{\sc i} column densities are large and in any case larger than 20.3.
Even though the models can fit the observations, little information can be extracted with confidence with this data quality. However, we noticed that the wind and disk models are more versatile than the spherical one and can be more easily adapted to the observations.

We show that more constraints could be obtained from better quality data with higher SNR and spectral resolution, especially if the Lyman-$\beta$ line can be observed. In particular, it seems possible to discriminate somehow between the three models.
Further observations with higher resolution are required to investigate these fascinating objects.

\section*{Acknowledgements}
In memoriam Hayley Finley.\par\noindent
We thank an anonymous referee for a thorough reading of the text and fruitful comments.\par\noindent
We thank the Physics and Astronomy department of Uppsala University in which part of the work presented here has been done.
\par\noindent
Funding for SDSS-III has been provided by the Alfred  P.  Sloan  Foundation,  the  Participating  Institutions, the National Science Foundation, and the U.S. Department of Energy Office of Science. The SDSS-III website is http://www.sdss3.org/. SDSS-III is managed by the Astrophysical Research Consortium for the Participating Institutions of the SDSS-III Collaboration including the University of Arizona,  the Brazilian Participation Group, Brookhaven National Laboratory, Carnegie Mellon  University,  University  of  Florida,  the  French Participation Group, the German Participation Group, Harvard University, the Instituto de Astrofisica de Canarias,  the  Michigan  State/Notre  Dame/JINA  Participation  Group,  Johns  Hopkins  University,  Lawrence Berkeley National Laboratory, Max Planck Institute for Astrophysics, Max Planck Institute for Extraterrestrial Physics, New Mexico State University, New York University, Ohio State University, Pennsylvania State University, University of Portsmouth, Princeton University, the Spanish Participation Group,  University of Tokyo, University of Utah, Vanderbilt University, University of Virginia, University of Washington, and Yale University.

\section*{Data availability}
Data used in this paper are from SDSS DR12 and are publicly available at https://dr12.sdss.org/.




\bibliographystyle{mnras}
\bibliography{main} 

\bsp	
\label{lastpage}
\end{document}